\documentclass[%
 reprint,
superscriptaddress,
 amsmath,amssymb,
 aps,
prb,
 floatfix,
]{revtex4-2}

\usepackage{placeins}
\usepackage{graphicx}%
\usepackage{dcolumn}%
\usepackage{bm}%

\usepackage{circuitikz}
\usepackage{subcaption}
\usepackage{balance}
\usepackage{tikz}

\def\cavity#1#2{ %
\draw (#1-2cm,#2-2cm) rectangle (#1+2cm,#2+2cm);
\draw (#1-2.5cm,#2-2.5cm) rectangle (#1+2.5cm,#2+2.5cm);}

\def\crystal#1#2#3{ %
\fill [color=blue!60] (#1-1cm,#2+1cm) to (#1-1cm,#2-1cm) to (#1+1cm,#2-1cm) to (#1+1cm,#2+1cm);
\draw (#1+0cm,#2+0cm) node [scale=1.8] {#3};}

\def\labelledbox#1#2#3#4#5{ %
\filldraw [fill=white] (#1-#3*0.5,#2+#4*0.5) rectangle (#1+#3*0.5,#2-#4*0.5);
\draw (#1,#2) node [scale=1.2] {#5};}

\def\attenuator#1#2#3{ %
\filldraw [fill=white] (#2-0.7,#3-0.5) rectangle (#2+0.7,#3);
\draw (#2,#3-0.25) node [scale=1.0] {#1 dB};}

\def\isolator#1#2{%
\filldraw [fill=white] (#1-0.4,#2-1) rectangle (#1+0.4,#2);
\draw [-{Triangle[scale=1.5]}] (#1,#2-0.9) to (#1,#2-0.1);}

\def\amplifier#1#2#3#4{%
\filldraw [fill=white,rotate around={#1:(#2,#3)}] (#2-0.4,#3) to (#2+0.4,#3) to (#2,#3+0.8) to (#2-0.4,#3);
\draw [rotate around={#1:(#2,#3)}] (#2,#3+0.3) node {#4};}

\def\supmagnet#1#2{%
\filldraw [fill=white] (#1-4.5,#2-2) rectangle (#1-3.5,#2+2);
\filldraw [fill=white] (#1+3.5,#2-2) rectangle (#1+4.5,#2+2);
\draw (#1-4,#2+1.5) circle [radius=0.23];
\draw (#1-4,#2+0.5) circle [radius=0.23];
\draw (#1-4,#2-0.5) circle [radius=0.23];
\draw (#1-4,#2-1.5) circle [radius=0.23];
\draw (#1+3.8,#2+1.7) to (#1+4.2,#2+1.3);
\draw (#1+4.2,#2+1.7) to (#1+3.8,#2+1.3);
\draw (#1+3.8,#2+0.7) to (#1+4.2,#2+0.3);
\draw (#1+4.2,#2+0.7) to (#1+3.8,#2+0.3);
\draw (#1+3.8,#2-0.7) to (#1+4.2,#2-0.3);
\draw (#1+4.2,#2-0.7) to (#1+3.8,#2-0.3);
\draw (#1+3.8,#2-1.7) to (#1+4.2,#2-1.3);
\draw (#1+4.2,#2-1.7) to (#1+3.8,#2-1.3);}

\begin{document}

\title{Identification of Paramagnetic Impurities in Calcium Fluoride \\ Doped with Various Inner Transition Metals}%

\author{Steven A. Samuels}
\affiliation{Quantum Technologies and Dark Matter Lab\\
The University of Western Australia, 35 Stirling Hwy, Crawley WA 6009}
\author{Michael Bartokos}
\affiliation{Vienna Center for Quantum Science and Technology, Atominstitut\\
 TU Wien, Stadionallee 2, 1020 Vienna, Austria}
\author{Kjeld Beeks}
\affiliation{Vienna Center for Quantum Science and Technology, Atominstitut\\
 TU Wien, Stadionallee 2, 1020 Vienna, Austria}
\author{William Campbell}
\affiliation{School of Mathematics and Physics\\
The University of Queensland, St Lucia QLD 4072}
\author{Aleksey Sadekov}
\affiliation{Centre for Microscopy, Characterisation and Analysis\\
The University of Western Australia, 35 Stirling Hwy, Crawley WA 6009}
\author{Dan Bevan}
\affiliation{Centre for Microscopy, Characterisation and Analysis\\
The University of Western Australia, 35 Stirling Hwy, Crawley WA 6009}
\author{Michael E. Tobar}
\affiliation{Quantum Technologies and Dark Matter Lab\\
The University of Western Australia, 35 Stirling Hwy, Crawley WA 6009}
\author{Maxim Goryachev}
\affiliation{Quantum Technologies and Dark Matter Lab\\
The University of Western Australia, 35 Stirling Hwy, Crawley WA 6009}

\date{\today}

\begin{abstract}
  Calcium fluoride (CaF$_{2}$) is the leading host for a solid-state $^{229}$Th nuclear clock and a host for rare-earth quantum technologies, and paramagnetic impurities at the parts-per-million level can limit both. We combine whispering-gallery-mode microwave spectroscopy at 20~mK with inductively coupled plasma mass spectrometry (ICPMS) to identify the paramagnetic impurities in undoped CaF$_{2}$ and in crystals doped with Th, Eu and Yb. Each interaction between a spin transition and a cavity mode is located by fitting the field dependence of the mode frequency and linewidth, which determines its crossing field with a typical uncertainty of 0.1~mT. In Th:CaF$_{2}$ the interactions reveal Mn$^{2+}$ through its resolved $^{55}$Mn hyperfine structure, although Mn is below the ICPMS detection limit of 0.15~ppm. A spin-Hamiltonian fit gives $g = 2.0023 \pm 0.0016$ and $A = (-95.2 \pm 0.6) \times 10^{-4}$~cm$^{-1}$; because only the lowest electronic level is populated at 20~mK, the sign of $A$ is also determined. In Eu:CaF$_{2}$, cubic Eu$^{2+}$ with literature parameters and Nd$^{3+}$ in a tetragonal site account for part of the interactions, and four further families of lines have the character of an $S$-state center with a zero-field splitting of order 10~GHz. In Yb:CaF$_{2}$ the interactions are assigned to Yb$^{3+}$ in cubic sites, and Yb$^{3+}$ dominates the microwave loss, lowering the mode quality factors at 20~mK. Cavity spectroscopy thus detects and identifies paramagnetic impurities below the detection limits of ICPMS, which alone cannot establish the impurity content relevant to these applications.
\end{abstract}

\maketitle

\section{Introduction}

Crystalline calcium fluoride (CaF$_2$) is a prototypical low-loss ionic dielectric that has long served as a model host material for precision optical and microwave spectroscopy. It is widely used as a scintillator and laser crystal~\cite{lucca2005yb,rodnyi2020physical}. Its wide band gap, simple cubic lattice, and exceptionally low intrinsic dielectric loss make it particularly attractive for experiments operating at cryogenic temperatures and high quality factors \cite{JACOB20032617,1320802}. In such regimes, the physical properties of CaF$_2$ are no longer limited by the host crystal itself, but instead are dominated by the presence of residual or intentionally introduced dilute impurities.

Dilute dopants in CaF$_2$, including rare-earth and actinide ions, play a dual role. Selected impurities such as Yb$^{3+}$ and Eu$^{3+}$ provide access to long-lived electronic and nuclear spin degrees of freedom, narrow optical transitions, and weak spin--phonon coupling, enabling applications in quantum memories, hybrid spin--photon systems, and precision spectroscopy \cite{liu2025long, kindem2018characterization, liu2003intensity, fossati2020frequency}. On the other hand, even parts-per-million concentrations of unintended paramagnetic impurities can introduce resonant microwave absorption, enhanced dielectric loss, and mode scattering in otherwise ultra-low-loss resonant structures \cite{harada2022controlling, thielemann2024optical}.

Actinide doping, particularly with thorium, has attracted renewed interest due to the unique low-energy nuclear transition in $^{229}$Th, which has been proposed and demonstrated to be the basis for a solid-state nuclear clock with unprecedented stability \cite{peik2021nuclear, kazakov2012performance,huang2026nuclear,toscani2026thorium}. CaF$_2$ has emerged as a leading host candidate for such studies~\cite{tiedau2024laser,ooi2026frequency,hiraki2025laser}, owing to its ability to incorporate thorium ions substitutionally while maintaining low lattice strain and minimal phonon-induced decoherence \cite{kazakov2012performance} resulting in narrow linewidths~\cite{morawetz2026continuous}. Other candidate hosts, such as ThF$_4$ and LiSAF, have yet to demonstrate narrow nuclear linewidths~\cite{elwell2024laser,zhang2024229thf4}. In this context, the identification of the thorium charge state, lattice site, and associated defect environment is essential, as these factors directly determine the coupling between nuclear, electronic, and electromagnetic degrees of freedom~\cite{hiraki2025laser}. Recently, Li et al.\ showed that the inhomogeneous nuclear linewidth in Th:CaF$_{2}$ is dominated by electric-field-gradient disorder, arising from lattice distortions around the Th dopants and, in the limit of zero doping, from defects of the host such as vacancies, interstitials and non-Th impurities~\cite{li2026origin}. Identifying the impurities in the host is therefore essential for building the most precise solid-state nuclear clock.

In this work, we investigate the role of dilute impurities in CaF$_2$ using high-sensitivity microwave spectroscopy at cryogenic temperatures and inductively coupled plasma mass spectrometry (ICPMS). By resolving and characterizing impurity-induced resonances, we identify the dominant paramagnetic species and quantify their coupling to confined microwave modes. Our results demonstrate that even extremely dilute impurities can govern the microwave response of nominally ultra-pure CaF$_2$, underscoring the necessity of impurity-level characterization in precision resonator and hybrid quantum experiments.

The paper is organized as follows. Section~\ref{sec:experiment} describes the crystals, the cavity spectroscopy setup, how spin--mode interactions are located in the measured transmission, and the ICPMS trace-element analysis. Section~\ref{sec:model} outlines the spin-Hamiltonian and mode-coupling models used to interpret the spectra. Section~\ref{sec:results} compares the dielectric loss of the crystals and then presents the spectroscopy of each: in Th-doped CaF$_{2}$ we identify Mn$^{2+}$; in Eu-doped CaF$_{2}$, Eu$^{2+}$ and Nd$^{3+}$ account for part of the interactions and the remainder are characterized; in Yb-doped CaF$_{2}$ the interactions are assigned to Yb$^{3+}$ in cubic sites. Section~\ref{sec:conclusion} summarizes the results, and Appendix~\ref{app:icpms} lists the full ICPMS results.

\section{Experiment}
\label{sec:experiment}

\begin{figure}[h]
  \centering
  \begin{subfigure}{0.2\textwidth}
      \centering
      \includegraphics[width=\linewidth]{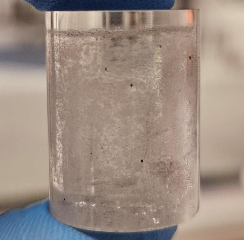}
      \caption{}
      \label{fig:ThCaF2_image}
  \end{subfigure}
  \hfill
  \begin{subfigure}{0.2\textwidth}
      \centering
      \includegraphics[width=\linewidth]{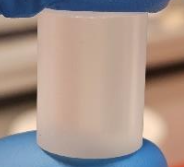}
      \caption{}
      \label{fig:CaF2_image}
  \end{subfigure}
  \hfill
  \begin{subfigure}{0.2\textwidth}
      \centering
      \includegraphics[width=\linewidth]{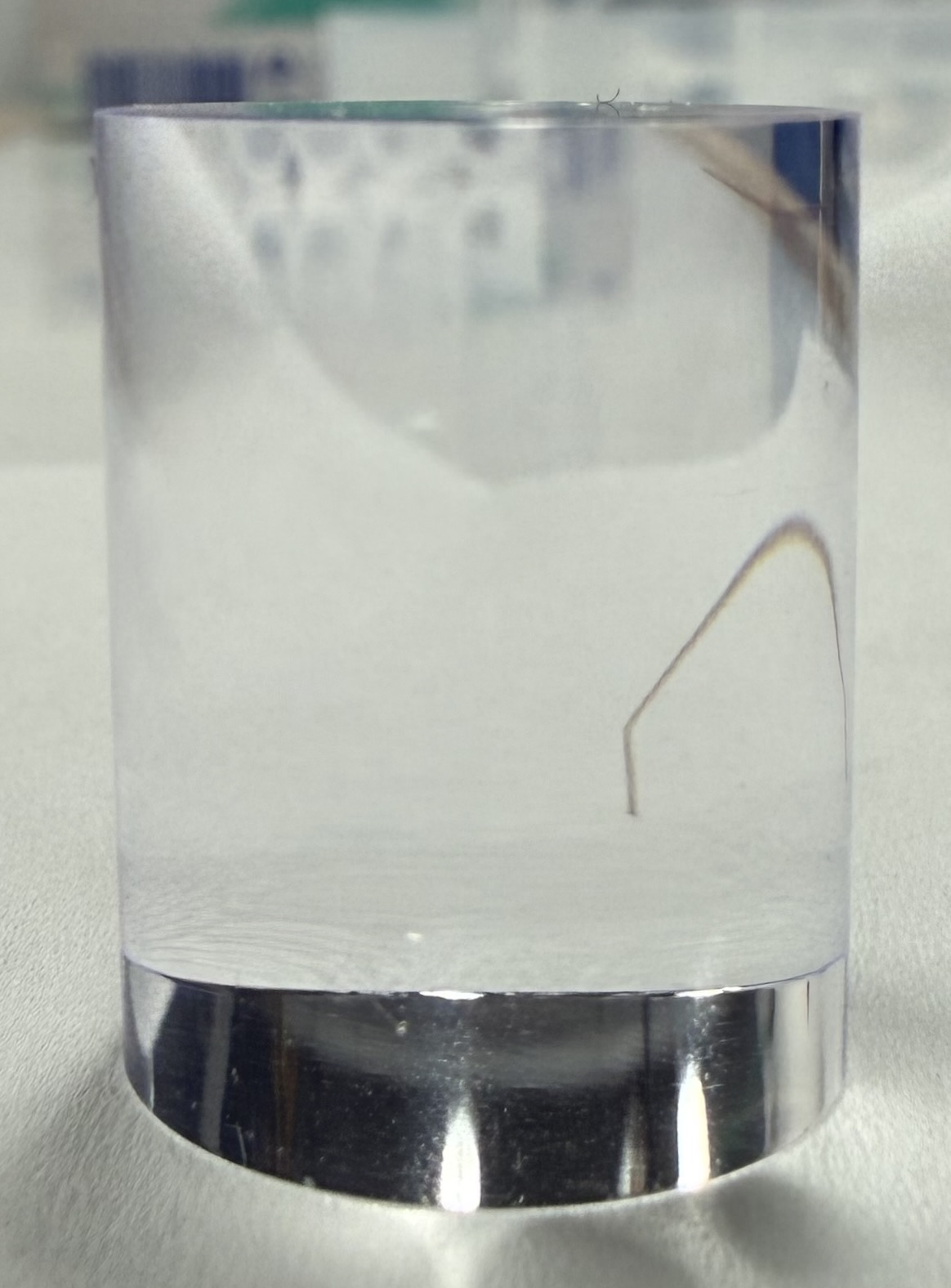}
      \caption{}
      \label{fig:EuCaF2_image}
  \end{subfigure}
  \caption{CaF\(_2\) samples used for microwave spectroscopy measurements: \(^{232}\)Th-doped (Fig.~\ref{fig:ThCaF2_image}), undoped (Fig.~\ref{fig:CaF2_image}), and Eu-doped (Fig.~\ref{fig:EuCaF2_image}).}
  \label{fig:CaF2_images}
\end{figure}

The identification of spin impurities is performed through analysis of the frequency and amplitude response of whispering-gallery modes (WGMs) sustained within a microwave cavity containing the crystal. In this approach, a cylindrical dielectric crystal is inserted into the cavity, where its low dielectric loss and strong photonic confinement enable high-quality-factor (high-$Q$) WGMs. These modes exhibit either transverse magnetic (WGH) or transverse electric (WGE) polarization with respect to the crystal’s cylindrical $z$-axis~\cite{giordano2023degeneracy}. A static external magnetic field is then applied and incrementally increased to high amplitudes. Electron paramagnetic resonance (EPR) occurs when the microwave photon energy matches the Zeeman splitting associated with impurity-ion spin transitions, leading to resonant absorption that promotes spins to higher energy states. Under dilute impurity concentrations, the coupling between the spin ensemble and the electromagnetic field perturbs the WGM spectrum, producing measurable frequency shifts and linewidth changes near the resonance field; only when the coupling exceeds the spin and mode linewidths does this become an avoided level crossing~\cite{PhysRevLett.110.157001}. Detection and analysis of these spectral features enable the identification of specific impurity species via their characteristic Zeeman signatures. Four CaF\(_2\) crystals were studied by microwave spectroscopy: undoped CaF\(_2\) (an IR-grade single crystal purchased from Korth GmbH), \(^{232}\)Th-doped CaF\(_2\) (grown from IR-grade CaF\(_2\) at TU Wien~\cite{schreitl2016growth}), Eu-doped CaF\(_2\) (Epic Crystal Co., Ltd.) and Yb-doped CaF\(_2\) (Stanford Advanced Materials). The first three are shown in Fig.~\ref{fig:CaF2_images}. A Tm-doped crystal was also analyzed by ICPMS (Table~\ref{tab:icpms}).

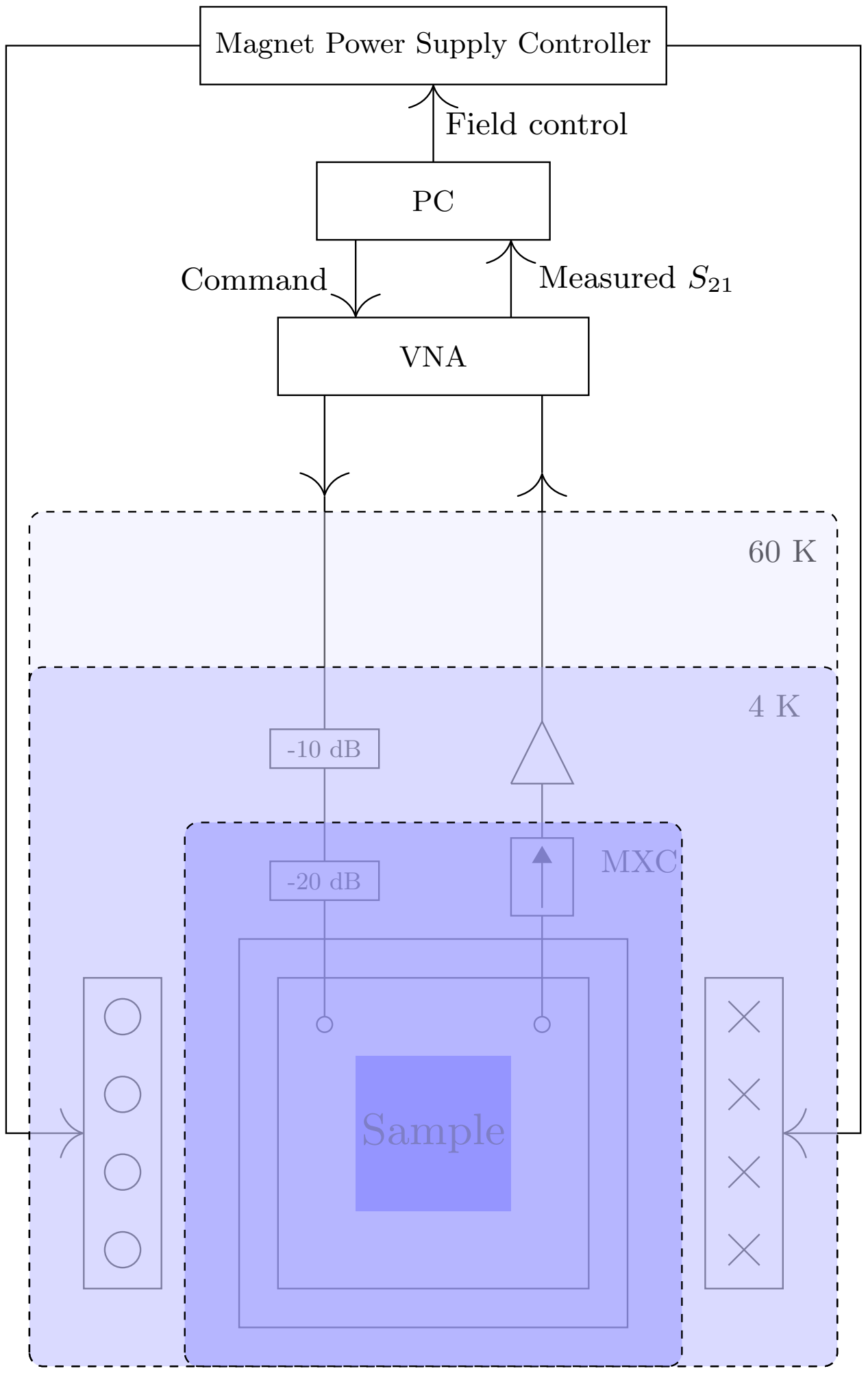
\begin{figure}[h]
  \begin{center}
    \scalebox{0.75}{\begin{tikzpicture}[line width=0.6pt]
        \draw [right] (2,3.5) node [scale=1.3] {MXC};
        \draw [right] (3.9,5.5) node [scale=1.3] {4~K};
        \draw [right] (3.9,7.5) node [scale=1.3] {60~K};
        \draw [right] (0,13) node [scale=1.3] {Field control};
        \draw [left] (-1.2,11) node [scale=1.3] {Command};
        \draw [right] (1.2,11) node [scale=1.3] {Measured $S_{21}$};
        \draw (-1.4,1.5) to (-1.4,8.2);
        \draw [-{>[scale=3]}] (-1.4,9.5) to (-1.4,8.2);
        \draw (-1.4,1.4) circle (1mm);
        \draw [-{>[scale=3]}] (1.4,1.5) to (1.4,8.5);
        \draw (1.4,8.5) to (1.4,9.5);
        \draw (1.4,1.4) circle (1mm);
        \draw [-{>[scale=3]}] (-3,14) to (-5.5,14) to (-5.5,0) to (-4.5,0);
        \draw [-{>[scale=3]}] (3,14) to (5.5,14) to (5.5,0) to (4.5,0);
        \cavity{0}{0}
        \crystal{0}{0}{Sample}
        \attenuator{-20}{-1.4}{3.5}
        \attenuator{-10}{-1.4}{5.2}
        \isolator{1.4}{3.8}
        \amplifier{0}{1.4}{4.5}{}
        \supmagnet{0}{0}
        \filldraw[fill=blue!10, fill opacity=0.4, draw=black, dashed, rounded corners=5pt] (-5.2,-3) -- (-5.2,8) -- (5.2,8) -- (5.2,-3) -- cycle;
        \filldraw[fill=blue!30, fill opacity=0.4, draw=black, dashed, rounded corners=5pt] (-5.2,-3) -- (-5.2,6) -- (5.2,6) -- (5.2,-3) -- cycle;
        \filldraw[fill=blue!50, fill opacity=0.4, draw=black, dashed, rounded corners=5pt] (-3.2,-3) -- (-3.2,4) -- (3.2,4) -- (3.2,-3) -- cycle;
        \labelledbox{0}{10}{4}{1}{VNA}
        \labelledbox{0}{12}{3}{1}{PC}
        \labelledbox{0}{14}{6}{1}{Magnet Power Supply Controller}
        \draw [-{>[scale=3]}] (-1,11.5) to (-1,10.5);
        \draw [-{>[scale=3]}] (1,10.5) to (1,11.5);
        \draw [-{>[scale=3]}] (0,12.5) to (0,13.5);
    \end{tikzpicture}}
  \end{center}
  \caption{Simplified schematic of the experiment. Microwave signals pass through thermal stages from room temperature (top) to the mixing chamber (MXC), where the crystal and cavity are located. The superconducting magnet applies a field along the crystal's $z$-axis. Temperature stages are indicated by blue-shaded regions, where darker shades represent colder stages.}
  \label{fig:CaF2_semiabstract_setup}
\end{figure}

As illustrated in Fig.~\ref{fig:CaF2_semiabstract_setup}, the crystal--cavity assembly is cooled to temperatures near 20~mK using a dilution refrigerator. Operation in this regime enables access to the exceptionally high quality factors achievable with WGM resonators at millikelvin temperatures \cite{farr2013ultrasensitive, hartman2024precision, hosain2018whispering, bourhill2019low}. Increased quality factors correspond to extended photon lifetimes within the resonator, thereby enhancing the signal-to-noise ratio for detecting spin resonance transitions. The effective coupling between WGMs and dilute spin impurities is further influenced by parameters such as impurity concentration, electromagnetic mode volume, and mode polarization \cite{PhysRevLett.110.157001}.

The microwave transmission response, $S_{21}(\omega)$, of the resonator WGMs is measured using a vector network analyzer (VNA) over an input power range from $-90$~dBm to $-30$~dBm. Excitation powers above this range are avoided to minimize unwanted crystal heating and to suppress nonlinear effects that could compromise the fidelity of EPR measurements. To reduce the contribution of thermal noise from room temperature electronics, cryogenic attenuators are installed on the input line at both the 4~K stage and the mixing chamber stage of the dilution refrigerator. On the output side, the cavity signal is routed through a microwave isolator before amplification by a low-noise cryogenic amplifier located at the 4~K stage. The isolator serves to block broadband noise emitted by the amplifier from propagating back into the microwave cavity.

Long-term frequency stability of the measurement setup is maintained by referencing the microwave sources and detection electronics to a hydrogen maser frequency standard.

Each spin--mode interaction (hereafter, interaction) is located from the field dependence of the WGM it perturbs. At every field step, the resonance frequency $f_{0}$ and half-width $\kappa$ of the mode are obtained from a fit to $S_{21}$; where only $|S_{21}|$ was recorded, $f_{0}$ is taken from a parabolic fit to the bottom of the resonance dip. A spin transition of half-width $\gamma$, detuned by $\Delta$ from the mode and coupled to it with strength $g_{\mathrm{c}}$, shifts and broadens the mode by $\delta f_{0} = g_{\mathrm{c}}^{2} \Delta / (\Delta^{2} + \gamma^{2})$ and $\delta\kappa = g_{\mathrm{c}}^{2} \gamma / (\Delta^{2} + \gamma^{2})$~\cite{PhysRevLett.110.157001}. Fitting these expressions, on a smooth background, to $f_{0}(B)$ and $\kappa(B)$ around each feature gives the center field of the interaction with its uncertainty, and its width.

\subsection{Trace-element analysis by ICPMS}

Impurity concentrations were measured by sector-field inductively coupled plasma mass spectrometry (ICPMS; Element XR) at the Centre for Microscopy, Characterisation and Analysis, University of Western Australia. A 70--113~mg piece of each crystal was crushed in an agate mortar, dissolved in boric acid--HCl and diluted gravimetrically in 2\% HNO$_3$ by an overall factor of $\approx 1.1\times10^{5}$. A procedural blank taken through the same chemistry was subtracted, and detection limits were set at three times the standard deviation of five instrument blanks. The transition metals, such as Ti, Cr, Mn, Fe, Co, Ni and Cu, were measured in medium resolution to separate them from isobaric interferences such as CaO$^{+}$ and CaF$^{+}$. Each element was normalized to Ca measured in the same solution, giving atomic fractions relative to Ca that do not depend on the dilution (Table~\ref{tab:icpms}; full results in Appendix~\ref{app:icpms}). ICPMS does not distinguish charge state or lattice site, so these values are upper bounds on the density of any single paramagnetic center, and contamination introduced during crushing is not captured by the blank.

\begin{table}[h]
  \caption{\label{tab:icpms} Dopants and paramagnetic impurities measured by ICPMS, in atomic ppm relative to Ca (1~ppm $= 2.45\times10^{16}$~cm$^{-3}$). Upper limits are detection limits, and values within a factor of 2 of the detection limit are marginal. Dashes mark signals dominated by carry-over from the preceding sample (Eu, Yb) or by Tm memory in the procedural blank. Co, Ni and V were below detection in every crystal ($<0.05$, $<0.3$ and $<0.07$~ppm).}
  \begin{ruledtabular}
  \begin{tabular}{lccccc}
  Ion & CaF$_2$ & Th:CaF$_2$ & Eu:CaF$_2$ & Yb:CaF$_2$ & Tm:CaF$_2$ \\
  \colrule
  Th & $<0.04$ & 5.9 & $<0.04$ & 0.11 & $<0.03$ \\
  Eu & $<0.02$ & $<0.02$ & 720 & -- & -- \\
  Yb & $<0.07$ & $<0.08$ & $<0.07$ & $9.0\times10^{3}$ & -- \\
  Tm & -- & -- & -- & -- & $6.5\times10^{3}$ \\
  \colrule
  Ti & 0.34 & 0.58 & 0.26 & 0.28 & 0.27 \\
  Cr & 0.33 & 0.16 & 0.12 & 0.73 & 0.16 \\
  Mn & 0.15 & $<0.15$ & $<0.14$ & $<0.20$ & 0.22 \\
  Fe & 1.1 & 1.4 & 1.8 & 2.4 & 1.0 \\
  Cu & 0.32 & 0.52 & $<0.16$ & 0.26 & 0.20 \\
  \colrule
  Ce & $<0.24$ & $<0.26$ & 0.28 & $<0.34$ & 0.35 \\
  Pr & $<0.06$ & $<0.06$ & 0.09 & $<0.08$ & $<0.05$ \\
  Nd & $<0.14$ & $<0.15$ & 0.52 & $<0.20$ & 0.15 \\
  Sm & $<0.05$ & $<0.05$ & 0.24 & $<0.07$ & 0.09 \\
  Gd & $<0.04$ & $<0.04$ & 0.27 & $<0.06$ & $<0.03$ \\
  Tb & $<0.02$ & $<0.02$ & 0.05 & $<0.03$ & 0.03 \\
  Dy & $<0.03$ & $<0.03$ & 0.23 & $<0.04$ & $<0.02$ \\
  Ho & $<0.02$ & $<0.02$ & 0.05 & 0.03 & 0.02 \\
  Er & $<0.02$ & $<0.02$ & 0.10 & 0.06 & 0.15 \\
  \end{tabular}
  \end{ruledtabular}
\end{table}

\section{Theoretical Model of the Experiment}
\label{sec:model}

The general Hamiltonian for a spin interacting with an external magnetic field is given by the following \cite{griffiths2019introduction}:

\begin{equation} \label{eq:genspinham}
  \hat{H} = \sum_{i} g_{i} \mu_{B} \mathbf{B}_{i} \cdot \mathbf{\hat{S}}_{i} + \sum_{i,j} \mathbf{J}_{ij} \mathbf{\hat{S}}_{i} \cdot \mathbf{\hat{S}}_{j}
\end{equation}

Here, $g_{i}$ is the $g$ factor and $\mathbf{B}_{i}$ the magnetic field at the site of the $i$th spin, $\mu_{B}$ is the Bohr magneton, $\mathbf{\hat{S}}_{i}$ is the spin operator for the $i$th spin, and $\mathbf{J}_{ij}$ is the exchange interaction between the $i$th and $j$th spins. EPR occurs when the energy difference between two eigenstates of (\ref{eq:genspinham}) matches the energy of a microwave photon from a WGM:

\begin{equation}\label{eq:resonance}
  h f_{\mathrm{WG}} = \Delta E = E_{m} - E_{n}
\end{equation}

Here, $f_{\mathrm{WG}}$ is the frequency of the WG mode and $\Delta E$ is the energy difference between the $m$th and $n$th eigenstates of (\ref{eq:genspinham}).

In EPR spectroscopy, the signal intensity is set by the net absorption strength. According to time-dependent perturbation theory, the net EPR absorption strength of a transition $\left| i \right\rangle \rightarrow \left| f \right\rangle$ is given by \cite{abragam2012electron}:

\begin{align}
  \rho_{i,f} &\propto \left| \bm{B}_{\text{WGM}} \cdot \left< f \left| \hat{\bm{\mu}} \right| i \right> \right|^{2} (p_{i} - p_{f}) \\ \nonumber
  \hat{\bm{\mu}} &= -g\mu_{B}\hat{\bm{S}}
\end{align}

Here, $\bm{B}_{\text{WGM}}$ denotes the magnetic field component of the WGM, and $\hat{\bm{\mu}}$ represents the magnetic dipole operator, while $p_{f}$ and $p_{i}$ denote the populations of the final and initial states, respectively. The external magnetic field defines the quantization axis so only components of the spin operator perpendicular to this axis, specifically $\hat{S}_{x}$ and $\hat{S}_{y}$, contribute to the transition probability. Transitions are therefore considered forbidden when the following condition is satisfied:

\begin{align} \label{eq:probability}
   \left| \left< f \right| \hat{S}_{x} \left| i \right> \right|^{2} + \left| \left< f \right| \hat{S}_{y} \left| i \right> \right|^{2} = 0
\end{align}

This condition can be used to filter out forbidden transitions from the simulated spectra.

\subsubsection{Magnetic Scatterers}

The coherent coupling of counter-propagating WGMs can be modeled using the following Hamiltonian \cite{PhysRevLett.99.173603}:

\begin{align} \label{eq:cp_WGMs}
    \hat{H} &= \hbar \omega_{k} \hat{a}^{\dagger}_{k} \hat{a}_{k} + \sum_{j} \hbar \omega_{j} \hat{a}^{\dagger}_{j} \hat{a}_{j} + \sum_{j} \hbar g_{kj} (\hat{a}^{\dagger}_{k} \hat{a}_{j} + \hat{a}^{\dagger}_{j} \hat{a}_{k})
\end{align}

Here, the free index $k$ represents the selected WGM and the summed index $j$ represents the WGMs with the opposite direction of propagation. As usual, $\hat{a}$ and $\hat{a}^{\dagger}$ are the annihilation and creation operators while $g_{kj}$ represents the coupling strength between the $k$th mode and the $j$th mode. For two counter-propagating modes, the solution can be reduced to the following:

\begin{align} \label{eq:mag_scatterer}
    \omega_{\pm} = \frac{\omega_{1} + \omega_{2}}{2} \pm \sqrt{g^{2} + \left( \frac{\omega_{1} - \omega_{2}}{2} \right)^{2}}
\end{align}

Here, $\omega_{\pm}$ are the normal-mode frequencies formed by the coupling between $\omega_{1}$ and $\omega_{2}$. For magnetic scatterers, the coupling strength will depend on the strength of the external magnetic field.

\section{Results and Discussion}
\label{sec:results}

\subsection{Dielectric Measurements}

\begin{figure}[h]
  \centering
  \begin{subfigure}{0.44\textwidth}
      \centering
      \includegraphics[width=\linewidth]{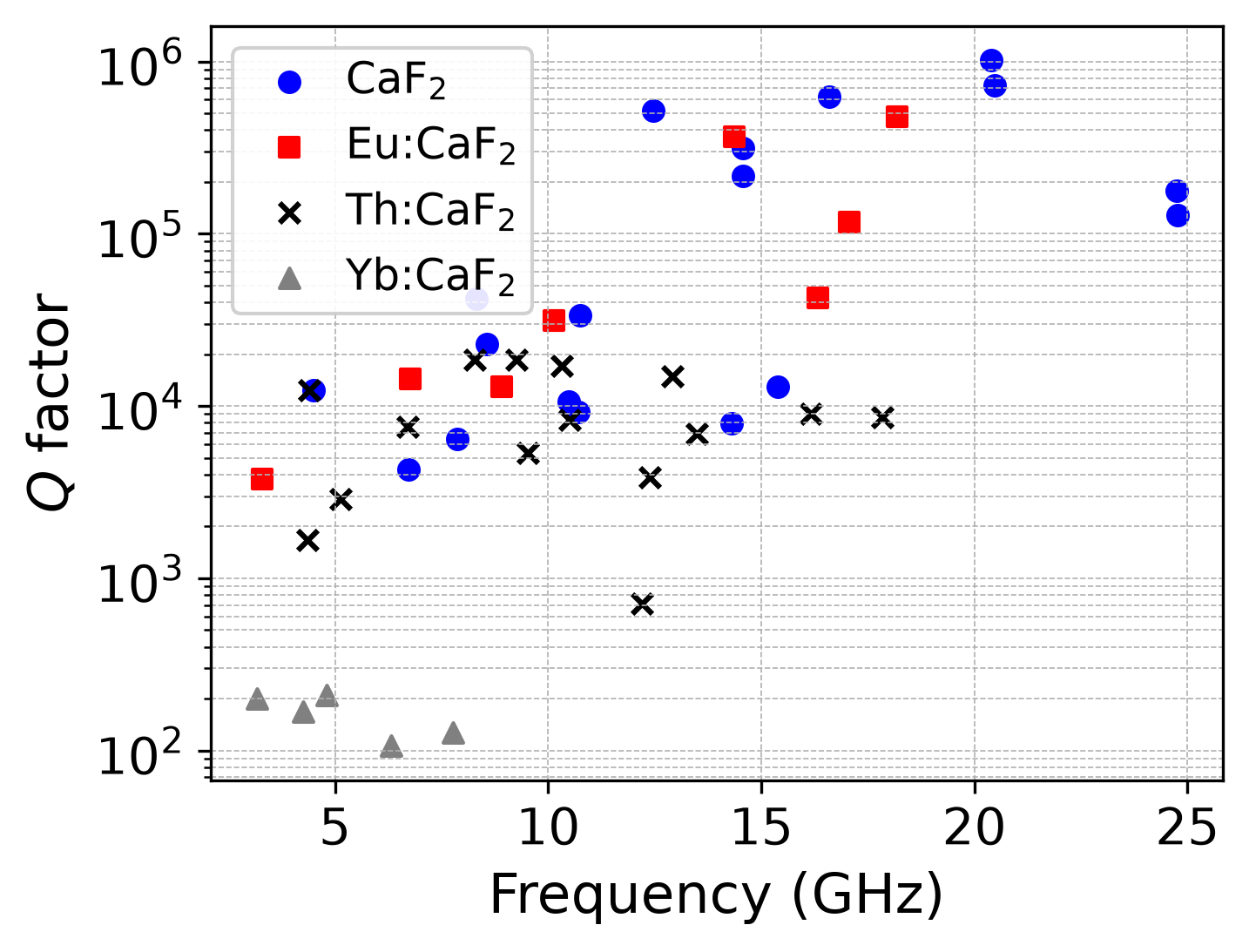}
      \caption{}
      \label{fig:Q_all_samples}
  \end{subfigure}
  \hfill
  \begin{subfigure}{0.44\textwidth}
      \centering
      \includegraphics[width=\linewidth]{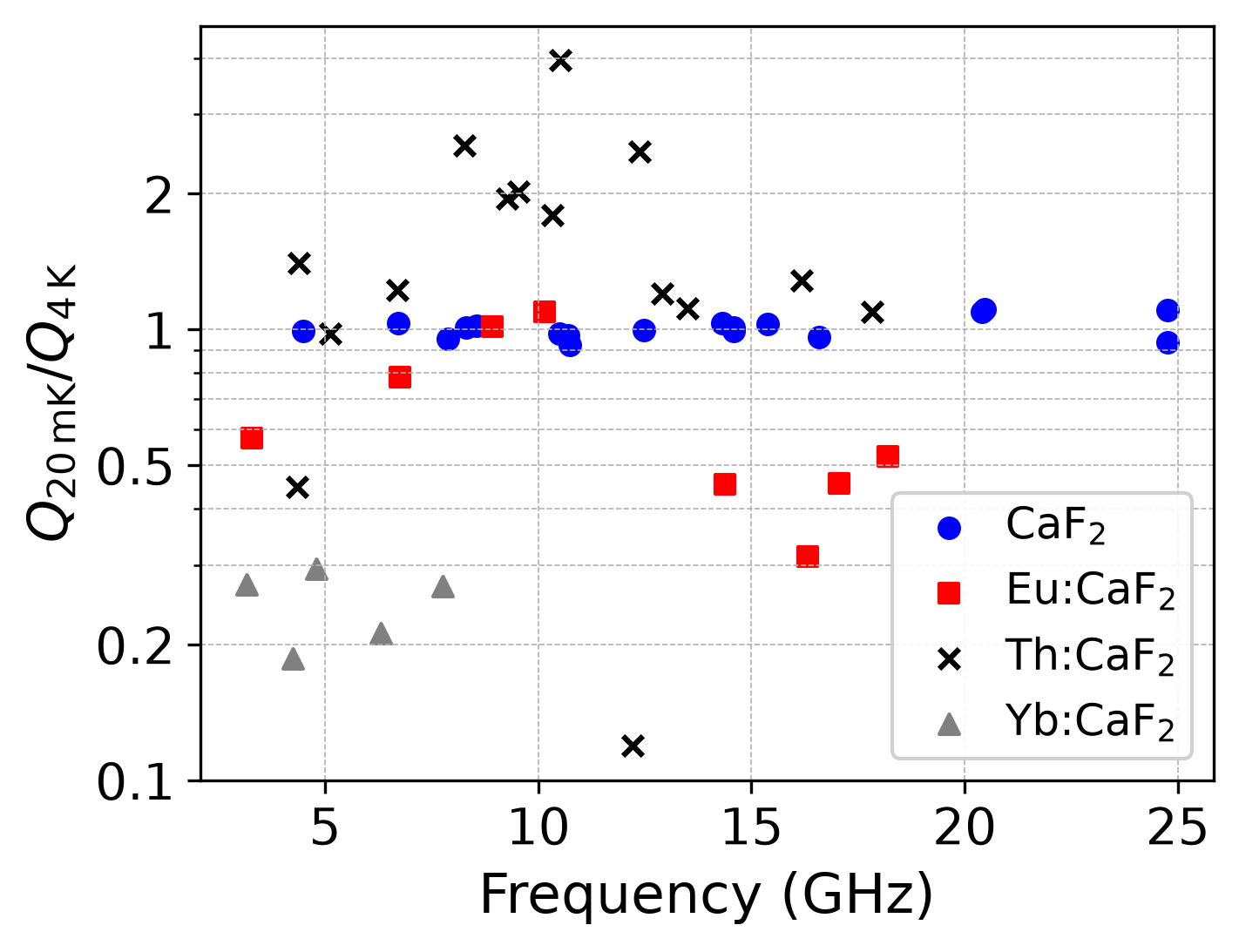}
      \caption{}
      \label{fig:q_ratio_all_samples}
  \end{subfigure}
  \caption{(a) $Q$ factors of microwave WGMs for all samples measured at 20~mK as a function of mode frequency. (b) Ratio of $Q$ factors at 20~mK to those at 4~K.}
  \label{fig:Qfactors}
\end{figure}

As shown in Fig.~\ref{fig:Q_all_samples}, the quality factors of the microwave modes employed for EPR spectroscopy are significantly higher in the undoped CaF$_2$ crystal and in the Eu-doped sample than in the Th-doped and Yb-doped crystals. The reduced $Q$ factors in the Th-doped sample are primarily attributed to its increased surface roughness, which enhances scattering losses and degrades the intrinsic quality factor of the whispering-gallery modes \cite{Rahachou2004}. The Yb-doped sample exhibits even lower $Q$ factors despite its smooth surface, implying a different loss mechanism. Yb$^{3+}$ (4$f^{13}$, $J=7/2$) is a strongly paramagnetic Kramers ion that occupies multiple crystallographic sites in CaF$_2$~\cite{PhysRev.155.279, kirton1969zeeman, Falin2004YbCaF2}, producing a broad distribution of EPR transitions and a substantial imaginary magnetic susceptibility $\chi''(\omega)$ across the microwave band~\cite{farr2013ultrasensitive}. This bulk spin-mediated loss exceeds the surface-scattering loss of the rougher Th-doped crystal. The high $Q$ of the Eu-doped sample is consistent with this picture: Eu$^{2+}$ is an $S$-state ion ($L=0$)~\cite{Baker_Bleaney_Hayes_1958, manenkov1956fine, ryterc1957} with narrow EPR linewidths; the residual loss is attributed to two-level system (TLS) absorption, which is weaker than the spin-mediated loss of the Yb-doped sample~\cite{creedon2011high, bourhill2019low}.

A systematic increase in $Q$ factor with increasing mode center frequency is observed for all samples. This trend is consistent with reduced surface participation ratios for higher-frequency whispering-gallery modes, as well as a weak frequency dependence of the dielectric loss tangent in CaF$_2$. Higher-order modes at elevated frequencies are also more strongly confined within the bulk of the crystal, thereby reducing sensitivity to surface imperfections.

As shown in Fig.~\ref{fig:q_ratio_all_samples}, the ratio of $Q$ factors at 20~mK to those at 4~K differs markedly between samples, revealing that the dominant loss mechanism is impurity-dependent rather than universal. The undoped CaF$_2$ shows negligible change between the two temperatures, indicating that the intrinsic loss of the host crystal is already temperature-saturated in this range. The Eu-doped and Yb-doped samples both exhibit reduced $Q$ factors at millikelvin temperatures relative to 4~K, with the Yb-doped sample showing the larger reduction in $Q$ on cooling. For the Eu-doped sample, the decrease is attributed to TLS absorption, which increases at lower temperatures as $\tanh(\hbar\omega/2k_BT) \to 1$~\cite{creedon2011high, bourhill2019low}. For the Yb-doped sample this is consistent with enhanced spin-mediated microwave absorption at lower temperatures: as $T$ decreases, the Boltzmann population difference across Yb$^{3+}$ spin transitions grows, increasing $\chi''(\omega)$ and the associated microwave loss~\cite{farr2013ultrasensitive}. The Th-doped sample shows the opposite behavior, with a significant improvement in $Q$ upon cooling. This is attributed to the suppression of phonon scattering on Th centers in the crystal lattice~\cite{JACOB20032617}, which constitutes the dominant loss channel at 4~K. As the thermal phonon population is frozen out at millikelvin temperatures, this scattering is strongly reduced, yielding the largest relative $Q$ improvement among all samples.

The observed variations in $Q$ factor have direct implications for the sensitivity of the EPR measurements. Higher $Q$ factors correspond to longer photon lifetimes and enhanced intracavity field amplitudes for a given input power, improving the signal-to-noise ratio and enabling the resolution of weak impurity-induced spectroscopic features. %

\subsection{Spectroscopy}

\subsubsection{Undoped CaF$_{2}$}

\begin{figure}[t]
    \centering
    \includegraphics[width=0.48\textwidth]{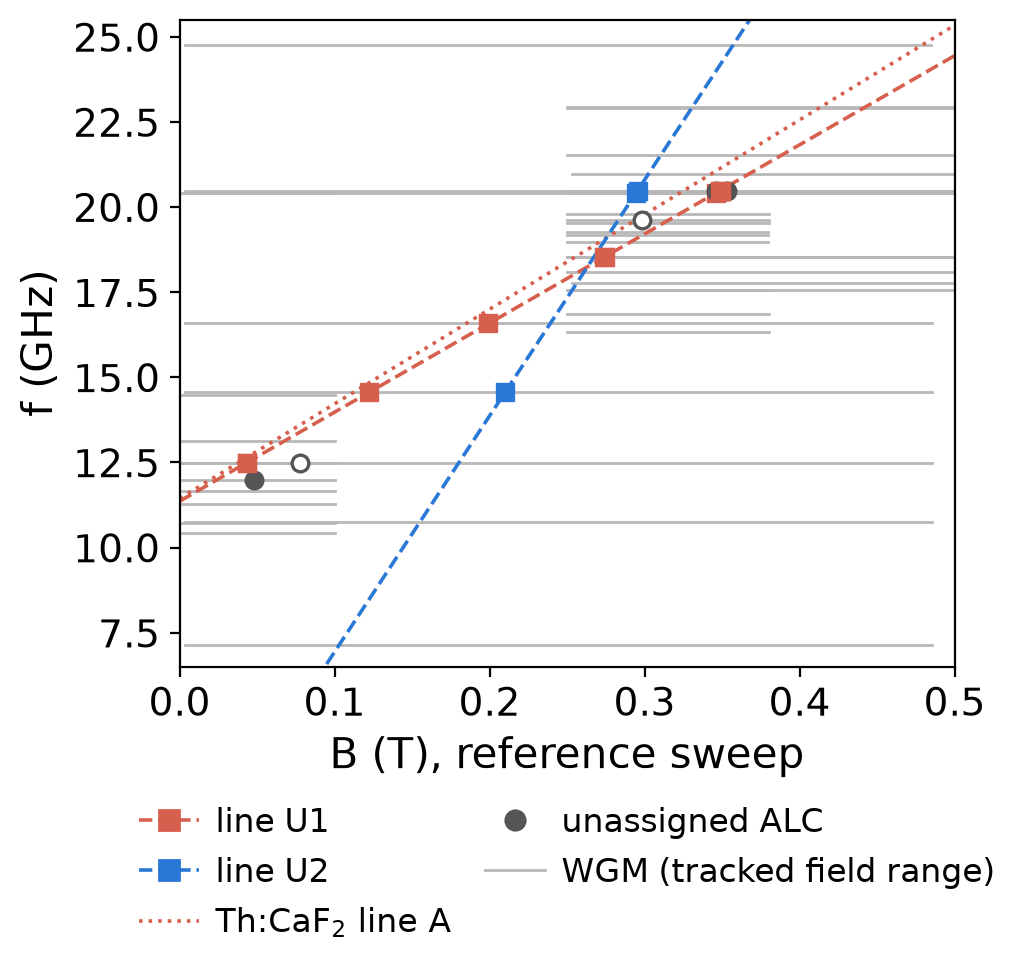}
    \caption{Spin--mode interactions observed in undoped CaF$_{2}$ at 20~mK, from six field sweeps in two cooldowns, with all fields shifted onto the scale of one sweep using the interactions they share. Gray horizontal lines: the WGMs over the field range in which each was tracked. Squares and dashed lines: interactions on the candidate lines U1 and U2, with straight-line fits (Table~\ref{tab:undoped_lines}). Dotted line: line A of Th:CaF$_{2}$, for comparison. Circles: interactions not assigned. Filled symbols: interactions found in two or more sweeps or above $5\sigma$; open symbols: weaker interactions. Horizontal error bars give the uncertainty in each interaction field, which is mostly smaller than the symbol.}
    \label{fig:undoped_CaF2_interactions}
\end{figure}

Interactions in the undoped crystal were extracted from six field sweeps in two cooldowns, covering 32 WGMs between 7.1 and 24.8~GHz at fields up to 0.5~T. The same interactions appear 3--4~mT lower in two of the sweeps than in the others, an offset in the magnetic field between sweeps. Each sweep is therefore shifted onto the field scale of one reference sweep using the interactions they share ($+3.8 \pm 1.0$ and $+3.4 \pm 0.9$~mT), and interactions found in several sweeps are merged. This gives 17 interactions, 15 of them found in two or more sweeps or above $5\sigma$ (Fig.~\ref{fig:undoped_CaF2_interactions}).

Most of the low-field features visible on d$S_{21}$/d$B$ color maps are not spin--mode interactions. In a sweep of the transmission from 10.7 to 12.7~GHz, $|S_{21}|$ at low field differs from its value at 0.1~T across the whole band, by up to $+3$~dB above about 11.3~GHz and $-2$~dB below it, and the difference fades smoothly by 40--80~mT. No part of this change moves with frequency as a spin transition would. A broadband change of the transmission background of this kind does not shift or broaden the modes as a spin--mode interaction does, and it is not analyzed further.

Nine interactions on six modes lie on a straight line, U1 (Table~\ref{tab:undoped_lines}), with $g = 1.871 \pm 0.006$ and a zero-field intercept of $+11.36$~GHz. It resembles line A of Th:CaF$_{2}$ ($g = 1.987$, $f_{0} = +11.44$~GHz; Sec.~\ref{sec:ThCaF2}). Both have the character of a fine-structure transition of an $S \geq 1$ center with a zero-field splitting of about 11.4~GHz, which would then be present in both crystals; the different slopes may reflect a different orientation of $\mathbf{B}$ relative to the center. Like line A, U1 is not exactly straight ($\chi^{2}/\nu = 178$). Three further interactions lie on a line through the origin, U2, with $g = 4.954 \pm 0.003$, but on only two distinct frequencies (14.6 and 20.4~GHz), so U2 is a weak candidate. Five interactions remain unassigned. ICPMS detects 0.15~ppm Mn in this crystal (Table~\ref{tab:icpms}), and the $g = 2$ field lies within the sweeps for the 7.1, 10.7 and 12.5~GHz modes, but no interactions on the Mn$^{2+}$ hyperfine lines are resolved.

\begin{table}[t]
  \caption{\label{tab:undoped_lines} Candidate lines through the interactions of undoped CaF$_{2}$ (Fig.~\ref{fig:undoped_CaF2_interactions}), all unassigned: $g$ (or the effective $g_{\mathrm{eff}}$) and zero-field intercept $f_{0}$ from a straight-line fit $f = g \mu_{B} B / h + f_{0}$, the number $N$ of interactions on each line and of modes they lie on, and the reduced $\chi^{2}$ of the fit. U2 is fitted through the origin. Members are the interactions within 1.5~mT of the line; uncertainties are scaled by $\sqrt{\chi^{2}/\nu}$. Line A of Th:CaF$_{2}$ (Sec.~\ref{sec:ThCaF2}), drawn in Fig.~\ref{fig:undoped_CaF2_interactions} for comparison, is listed below the rule.}
  \begin{ruledtabular}
  \begin{tabular}{lccccc}
  Line & $g$ & $f_{0}$ (GHz) & $N$ & Modes & $\chi^{2}/\nu$ \\
  \colrule
  U1 & $1.871 \pm 0.006$ & $+11.36 \pm 0.02$ & 9 & 6 & 178 \\
  U2 & $4.954 \pm 0.003$ & 0 & 3 & 3 & 14 \\
  \colrule
  A (Th) & $1.987 \pm 0.008$ & $+11.44 \pm 0.02$ & 4 & 4 & 54 \\
  \end{tabular}
  \end{ruledtabular}
\end{table}

\subsubsection{Th:CaF$_{2}$}
\label{sec:ThCaF2}

\begin{figure*}[t]
    \centering
    \includegraphics[width=0.9\linewidth]{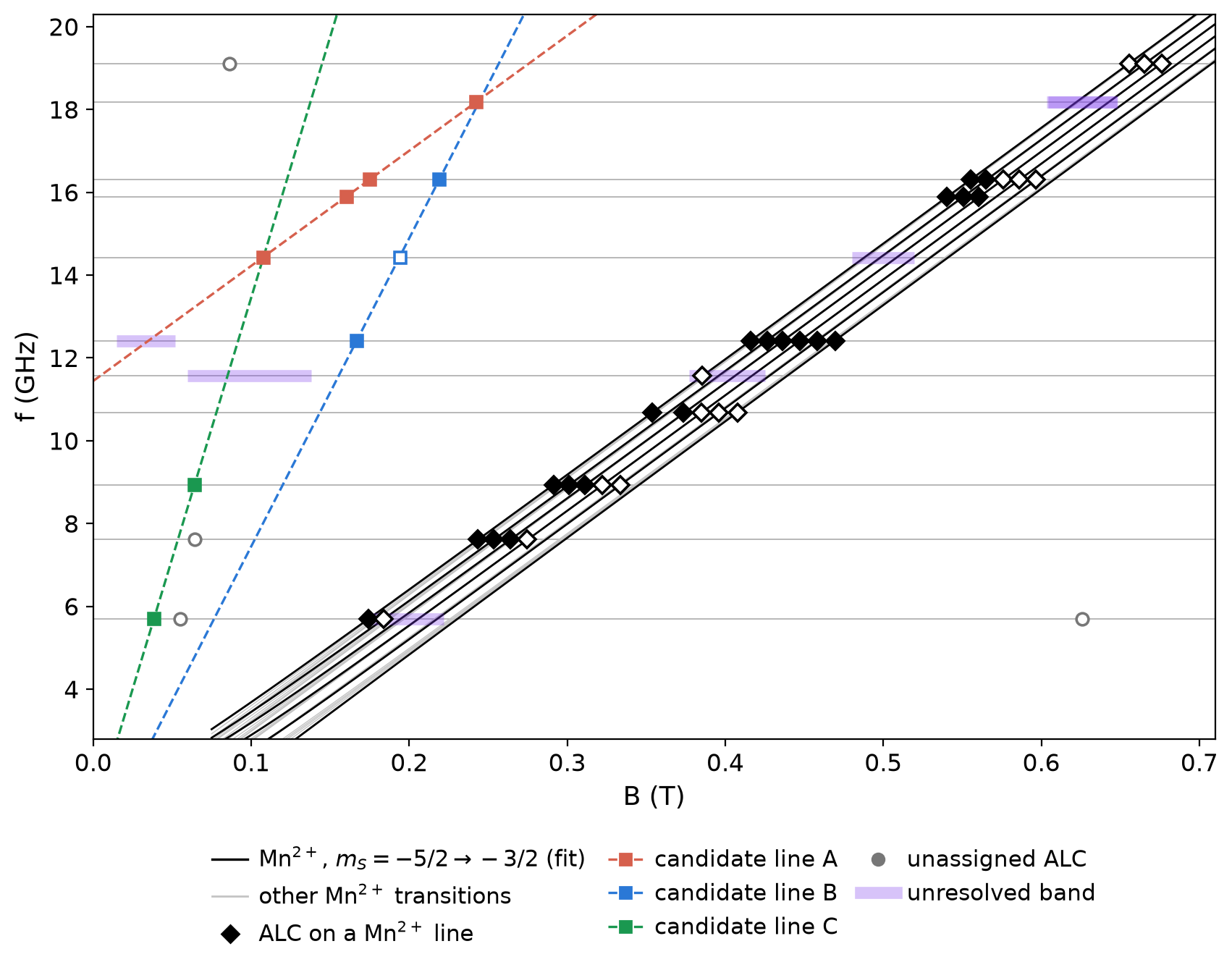}
    \caption{Spin--mode interactions observed in Th:CaF$_{2}$ at 20~mK, compared with the $^{55}$Mn$^{2+}$ spectrum fitted to them (Table~\ref{tab:Mn2pl_params}). Horizontal gray lines mark the WGM frequencies. Black lines: the $m_{S} = -5/2 \to -3/2$ transition of each hyperfine line from Eq.~(\ref{eq:Mn2pl_ham}) with $\mathbf{B}$ along [001], the Mn$^{2+}$ transitions originating from the populated $m_{S} = -5/2$ manifold at 20~mK; light gray lines: the other Mn$^{2+}$ transitions. Black diamonds: interactions on the Mn$^{2+}$ hyperfine lines. Colored squares and dashed lines: interactions on the candidate lines A--C, with straight-line fits (Table~\ref{tab:ThCaF2_lines}). Gray circles: interactions not assigned. Purple bars: broad features (half-width above 8~mT) not resolved into separate interactions, drawn as center $\pm$ two half-widths. Filled symbols: interactions found in both sweep directions or, for modes measured in one direction, above $5\sigma$; open symbols: weaker interactions ($3$--$5\sigma$, or found in only one of two sweeps). Horizontal error bars give the uncertainty in each interaction field, which is mostly smaller than the symbol.}
    \label{fig:ThCaF2_interactions}
\end{figure*}

\begin{table}[t]
  \caption{\label{tab:Mn2pl_params} Spin-Hamiltonian parameters of Mn$^{2+}$ in CaF$_{2}$, Eq.~(\ref{eq:Mn2pl_ham}). This work: $g$ and $A$ fitted to the fields of 20 resolved interactions on the hyperfine lines, on seven modes (Fig.~\ref{fig:ThCaF2_interactions}), each taken as the $m_{S} = -5/2 \to -3/2$ transition, with $\mathbf{B}$ along [001] and $a$ fixed at the literature value; uncertainties combine the statistical error, scaled by $\sqrt{\chi^{2}/\nu} \approx 2.1$, with the largest change when the interactions of one mode are left out. The fitted $g$ also absorbs any error in the field calibration of the magnet. The sign of $A$ is fixed by the population of only the lowest level at 20~mK (see text); in the literature only the relative sign of $a$ and $A$ is determined~\cite{Baker_Bleaney_Hayes_1958}, and signs are as quoted in each reference.}
  \begin{ruledtabular}
  \begin{tabular}{lccc}
  Source & $g$ & $a$ ($10^{-4}$~cm$^{-1}$) & $A$ ($10^{-4}$~cm$^{-1}$) \\
  \colrule
  This work & $2.0023 \pm 0.0016$ & $0.6$ (fixed) & $-95.2 \pm 0.6$ \\
  Ref.~\cite{Baker_Bleaney_Hayes_1958} & $1.998 \pm 0.003$ & $0.6 \pm 0.4$ & $-97.8 \pm 0.1$ \\
  Ref.~\cite{VANGORKOM1970905} & $2.002 \pm 0.001$ & $0.6$ & $95.2 \pm 0.1$ \\
  \end{tabular}
  \end{ruledtabular}
\end{table}

\begin{table}[t]
  \caption{\label{tab:ThCaF2_lines} Candidate lines through the interactions of Fig.~\ref{fig:ThCaF2_interactions} that do not lie on a Mn$^{2+}$ transition: $g$ (or the effective $g_{\mathrm{eff}}$) and zero-field intercept $f_{0}$ from a straight-line fit $f = g \mu_{B} B / h + f_{0}$, and the number $N$ of interactions on each line. Line B is fitted through the origin. Uncertainties are scaled by $\sqrt{\chi^{2}/\nu}$ where this exceeds 1.}
  \begin{ruledtabular}
  \begin{tabular}{llccc}
  Line & Assignment & $g$ & $f_{0}$ (GHz) & $N$ \\
  \colrule
  A & unassigned & $1.987 \pm 0.008$ & $+11.44 \pm 0.02$ & 4 \\
  B & unassigned & $5.318 \pm 0.005$ & 0 & 3 \\
  C & unassigned & $9.00 \pm 0.02$ & $+0.85 \pm 0.03$ & 3 \\
  \end{tabular}
  \end{ruledtabular}
\end{table}

\begin{figure}[h]
    \centering
    \includegraphics[width=0.99\linewidth]{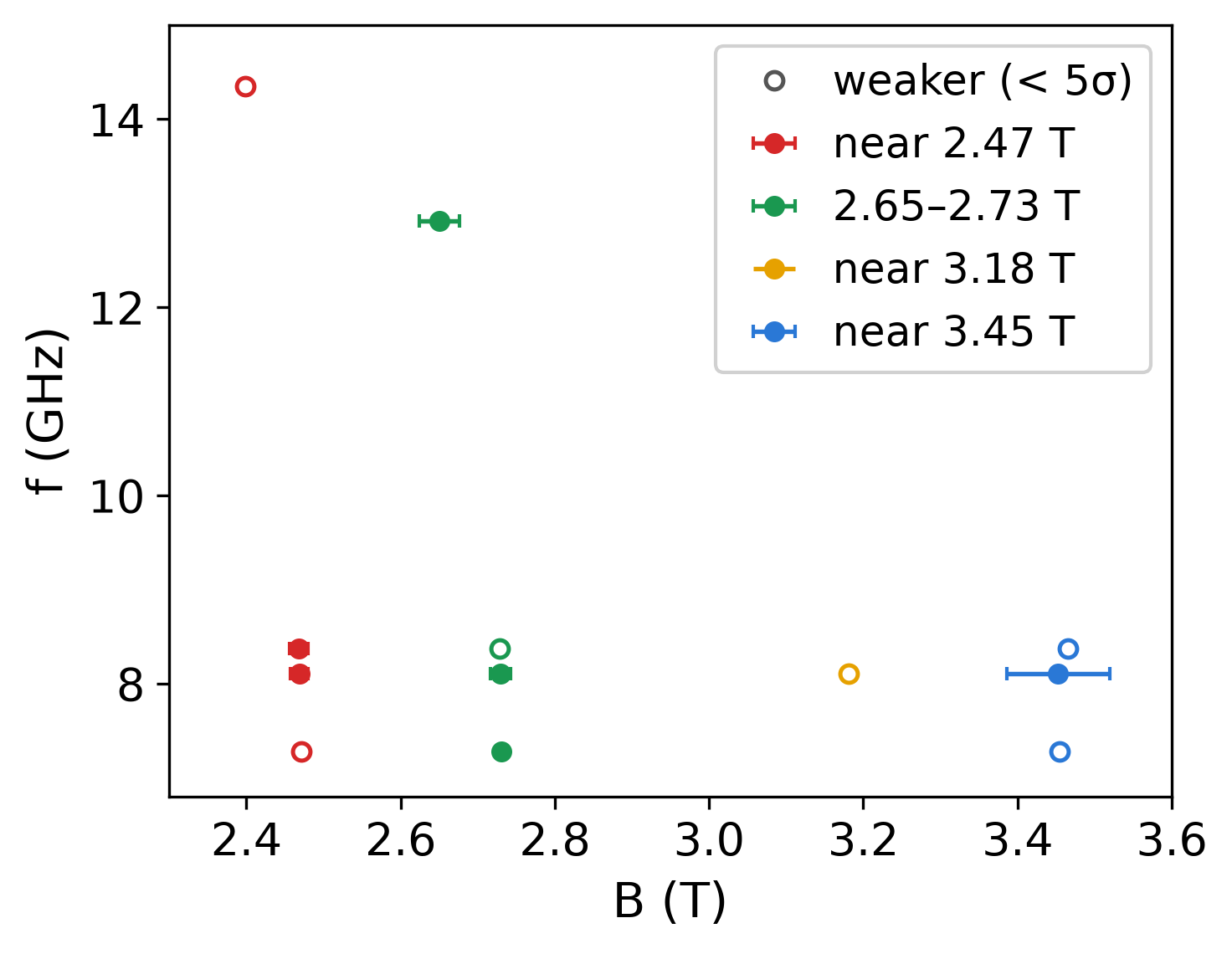}
    \caption{Spin--mode interactions observed in Th:CaF$_{2}$ at 4~K at high magnetic field. Each point is the center field $B_{0}$ of an interaction, fitted to the field dependence of the mode frequency; for filled symbols its uncertainty (at most 5~mT) is smaller than the symbol. Filled symbols: interactions detected at more than $5\sigma$, with the fitted half-width as the horizontal bar. Open symbols: weaker features. Colors group interactions at similar fields. Interactions on the 10.66, 11.52 and 11.59~GHz modes, where the mode frequency could not be tracked, are not shown.}
    \label{fig:ThCaF2_high_B}
\end{figure}

The features observed in this sample are of four types: interactions on the hyperfine lines of Mn$^{2+}$, interactions at low field that do not lie on a Mn$^{2+}$ transition (both in Fig.~\ref{fig:ThCaF2_interactions}), magnetic scattering of microwave modes (Fig.~\ref{fig:scatter_cplot_9p53ghz}), and unassigned interactions at high magnetic field (Fig.~\ref{fig:ThCaF2_high_B}). The Mn$^{2+}$ center is modeled by the following:

\begin{figure}[h]
    \centering
    \includegraphics[width=0.95\linewidth]{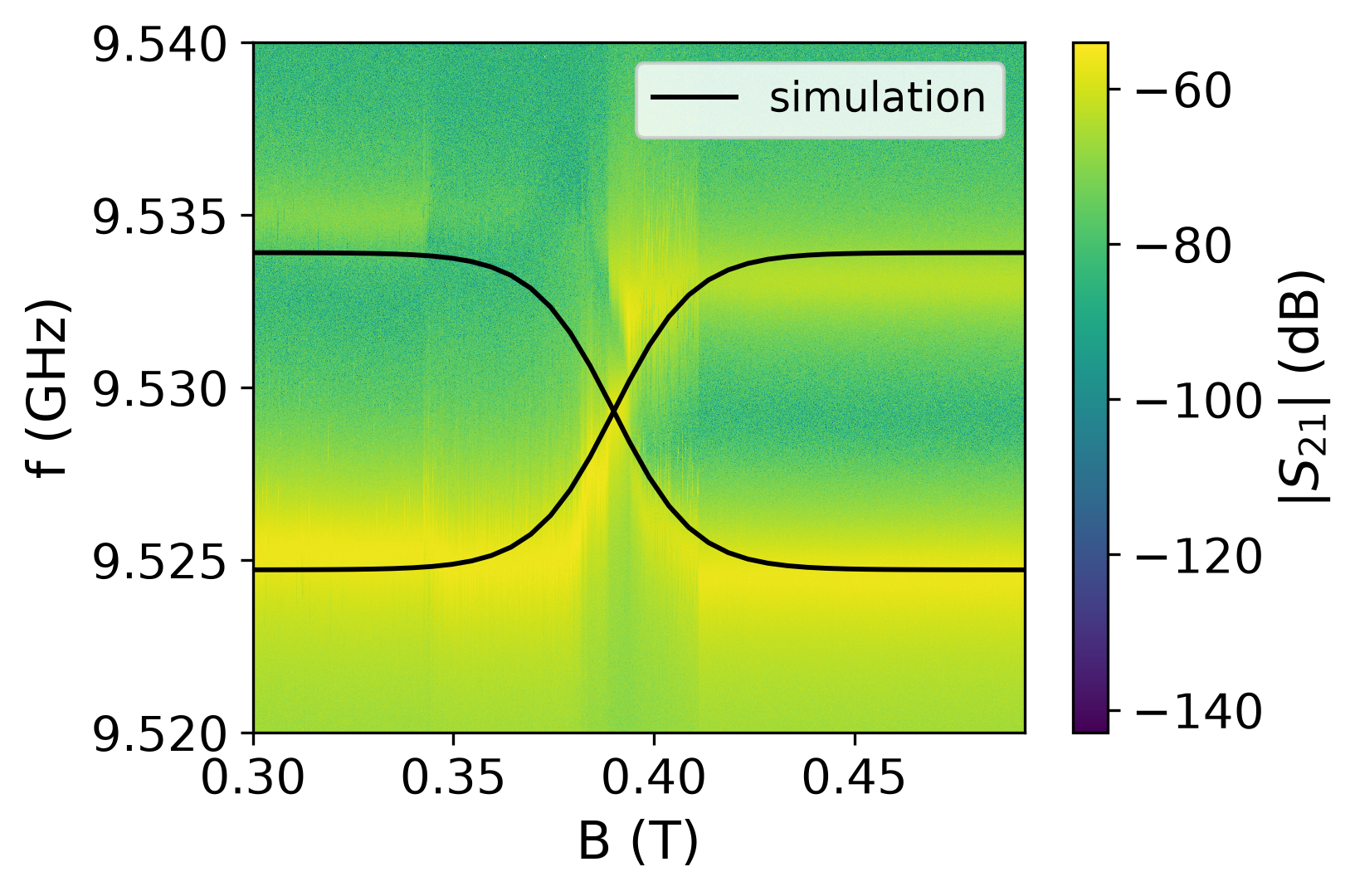}
    \caption{Scattering interaction in the doublet mode centered around 9.53 GHz in Th:CaF$_{2}$. A magnetic scatterer simulation using Eq.~(\ref{eq:mag_scatterer}) is overlaid on the plot.}
    \label{fig:scatter_cplot_9p53ghz}
\end{figure}

\begin{align} \label{eq:Mn2pl_ham}
    \hat{H} &= g \mu_{B} \mathbf{B} \cdot \mathbf{S} \nonumber \\
    &+ \frac{1}{6} a \left[ S_{x}^{4} + S_{y}^{4} + S_{z}^{4} - \frac{1}{5} S(S+1)(3S^{2} + 3S - 1) \right] \nonumber \\
    &+ A (\mathbf{S} \cdot \mathbf{I})
\end{align}

The assignment rests on the hyperfine structure. Near $g = 2$ the interactions of each mode form the sextet expected for $^{55}$Mn ($I = 5/2$, 100\% abundant): at 12.4~GHz all six lines are resolved, 10.0--11.2~mT apart, and one to three lines are resolved on each of six further modes between 5.7 and 16.3~GHz (Fig.~\ref{fig:ThCaF2_interactions}). Lines this closely spaced overlap, so the interactions near $g = 2$ on each mode were fitted together in a single joint fit. At 20~mK nearly all Mn$^{2+}$ ions are in the $m_{S} = -5/2$ level, so each interaction is the $m_{S} = -5/2 \to -3/2$ transition of one hyperfine line; with the literature parameters, this assignment fits the interaction fields far better than the central $m_{S} = -1/2 \to +1/2$ transition. Three weaker interactions on the 19.1~GHz mode (open diamonds in Fig.~\ref{fig:ThCaF2_interactions}) also lie in its sextet region, but each was found in only one sweep at $3$--$5\sigma$, and they are not used in the fit. Fitting $g$ and $A$ of Eq.~(\ref{eq:Mn2pl_ham}) to the 20 resolved lines gives the values in Table~\ref{tab:Mn2pl_params}, with $\chi^{2}/\nu = 4.5$. The cubic term $a$ cannot be separated from $g$ by these data and is fixed at the literature value. The fitted $|A|$ agrees with Ref.~\cite{VANGORKOM1970905} and is $2.6 \times 10^{-4}$~cm$^{-1}$ smaller than in Ref.~\cite{Baker_Bleaney_Hayes_1958}. Because only the lowest level is populated, the sign of $A$ is also determined: the second-order hyperfine shifts of the $m_{S} = -5/2 \to -3/2$ transition depend on it, and $A > 0$ fits far worse ($\chi^{2}/\nu = 43$). Mn was not detected by ICPMS in this crystal ($<0.15$~ppm, Table~\ref{tab:icpms}), which places the Mn$^{2+}$ density below $3.7 \times 10^{15}$~cm$^{-3}$.

Thirteen interactions, nine of them found in both sweep directions or at more than $5\sigma$, do not lie on a Mn$^{2+}$ transition. We tested every straight line through two of these interactions on different modes for further interactions within $3\sigma$ (at least 0.5~mT) on other modes. Three candidate lines result (Table~\ref{tab:ThCaF2_lines}). Line A passes through four interactions on four modes, with $g = 1.987$ and an intercept of $+11.4$~GHz, the behavior of a fine-structure transition of an $S \geq 1$ center with a zero-field splitting of order 10~GHz. With the interactions placed at random fields, a line through four interactions appears in 0.6\% of trials. Line A is not exactly straight ($\chi^{2}/\nu = 54$). Lines B and C pass through three interactions each; lines through three interactions appear in 38\% of the random trials, and each crosses 14 modes with interactions on only three or four of them, so they remain candidates. Line B passes through the origin with $g = 5.318$, as for an effective spin-1/2 Kramers doublet; this value lies within the range of $g$ of Er$^{3+}$ in one of its tetragonal sites and in its trigonal sites, and close to one of the $g$ values of Dy$^{3+}$ in cubic sites~\cite{weber1964paramagnetic}. Rare earths were below the ICPMS detection limits in this crystal (Table~\ref{tab:icpms}), but the strong Mn$^{2+}$ interactions show that such limits do not exclude them. Four interactions remain unassigned. Six broad features are not resolved into separate interactions; four of them lie in the Mn$^{2+}$ sextet region of their mode and are probably unresolved hyperfine lines.

Magnetic scattering of microwave modes was also observed at fields of roughly 0.35 to 0.41~T. Figure~\ref{fig:scatter_cplot_9p53ghz} shows an example of such an interaction on a magnetic-field-dependent spectrum color plot.

\begin{figure*}[t]
    \centering
    \includegraphics[width=0.9\linewidth]{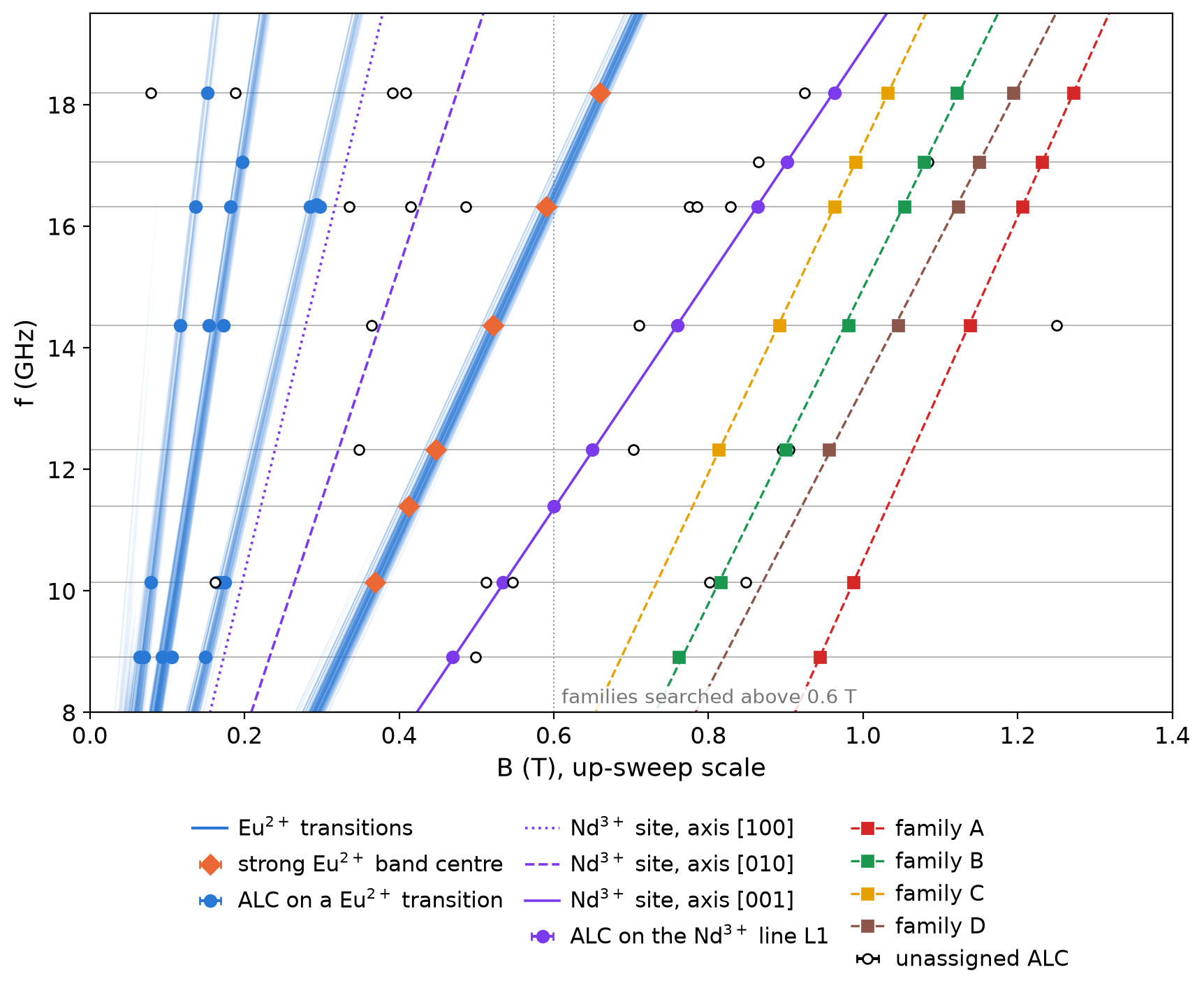}
    \caption{Spin--mode interactions observed in Eu:CaF$_{2}$, compared with the literature spectra of Eu$^{2+}$~\cite{ryterc1957} and Nd$^{3+}$~\cite{bleaney1956uranium} calculated for a single direction of $\mathbf{B}$, with no spin-Hamiltonian parameters adjusted (Table~\ref{tab:eu2pl_params}). Horizontal gray lines mark the WGM frequencies. Blue lines: Eu$^{2+}$ transitions from Eq.~(\ref{eq:Eu2pl_ham}) for both isotopes, shaded on a logarithmic scale spanning three decades by the population difference at 20~mK times the squared transition matrix element. Purple lines: the single line of each of the three tetragonal Nd$^{3+}$ sites, with the site axis along [100] (dotted), [010] (dashed) or [001] (solid). Orange diamonds: centers of the strong Eu$^{2+}$ band. Blue circles: interactions within $3\sigma$ of a populated Eu$^{2+}$ transition. Purple circles: interactions on the [001] Nd$^{3+}$ line (L1). Colored squares and dashed lines: interactions in families A--D, each lying on a common straight line, with straight-line fits (Table~\ref{tab:EuCaF2_g}); families were searched for only above 0.6~T (dotted vertical line). Open circles: interactions not assigned. Horizontal error bars give the uncertainty in each interaction field, which is mostly smaller than the symbol. Fields measured on down-sweeps are shifted onto the up-sweep scale to correct for magnet lag.}
    \label{fig:EuCaF2_Eu2pl_sim}
\end{figure*}

In a separate measurement at 4~K with fields up to 4.2~T, further interactions were found between 2.4 and 3.5~T (Fig.~\ref{fig:ThCaF2_high_B}). These lie far above the fields at which a $g \approx 2$ transition crosses these modes (0.26--0.68~T). An $S$-state ion with a small zero-field splitting, such as Eu$^{2+}$, Gd$^{3+}$ or Mn$^{2+}$ in a cubic site, has no transitions at 7--15~GHz in this field range, so these interactions are not due to the Mn$^{2+}$ center identified above. The interactions near 2.47 and 2.73~T occur at the same field, within 3~mT, in the 7.28, 8.10 and 8.37~GHz modes: a spin line through either set would have to change in frequency by more than 120 and 240~GHz/T, respectively. Their half-widths are 2--13~mT. The origin of these features remains unidentified.

\subsubsection{Eu:CaF$_{2}$}

The following Hamiltonian is used to simulate Eu$^{2+}$ in CaF$_{2}$, in the notation of Ref.~\cite{Baker_Bleaney_Hayes_1958}, where $b_{4} = 60 B_{4}$ and $b_{6} = 1260 B_{6}$:

\begin{align} \label{eq:Eu2pl_ham}
    \hat{H} &= g \mu_{B} \mathbf{B} \cdot \mathbf{\hat{S}} + \frac{b_{4}}{60}(\hat{O}_{4}^{0} + 5 \hat{O}_{4}^{4}) \\ \nonumber
    &+ \frac{b_{6}}{1260}(\hat{O}_{6}^{0} - 21 \hat{O}_{6}^{4}) + A \mathbf{\hat{S}} \cdot \mathbf{\hat{I}}
\end{align}

The first term represents the Zeeman interaction, the second and third terms represent the cubic crystal-field (zero-field) splitting in terms of the Stevens operators, and the last term represents the hyperfine interaction ($S = 7/2$, $I = 5/2$ for both $^{151}$Eu and $^{153}$Eu). The parameters $g$, $b_{4}$, $b_{6}$ and $A$ are taken from the EPR literature without adjustment (Table~\ref{tab:eu2pl_params}).

\begin{table}[t]
  \caption{\label{tab:eu2pl_params} Spin-Hamiltonian parameters used to simulate the spectra in Fig.~\ref{fig:EuCaF2_Eu2pl_sim}; none were adjusted. Eu$^{2+}$: Eq.~(\ref{eq:Eu2pl_ham}) with the values of Ryter~\cite{ryterc1957} and the uncertainties quoted in Ref.~\cite{Baker_Bleaney_Hayes_1958}; only relative signs are known ($b_{4}$ and $A$ have the same sign, $b_{6}$ the opposite). Nd$^{3+}$: axial $g$-tensor of the tetragonal site~\cite{bleaney1956uranium}. The direction of $\mathbf{B}$ is given relative to [110]: the rotation is fitted to the strong Eu$^{2+}$ band centers and the tilt to the $g$-value of L1 (Table~\ref{tab:EuCaF2_g}).}
  \begin{ruledtabular}
  \begin{tabular}{lc}
  Parameter & Value \\
  \colrule
  \multicolumn{2}{l}{Eu$^{2+}$, cubic site ($S = 7/2$, $I = 5/2$)} \\
  $g$ & $1.9927 \pm 0.0010$ \\
  $b_{4}$ ($10^{-4}$~cm$^{-1}$) & $-55.75 \pm 0.28$ \\
  $b_{6}$ ($10^{-4}$~cm$^{-1}$) & $+0.25 \pm 0.06$ \\
  $A(^{151}\mathrm{Eu})$ ($10^{-4}$~cm$^{-1}$) & $-34.07 \pm 0.10$ \\
  $A(^{153}\mathrm{Eu})$ ($10^{-4}$~cm$^{-1}$) & $-15.10 \pm 0.10$ \\
  \colrule
  \multicolumn{2}{l}{Nd$^{3+}$, tetragonal site ($S' = 1/2$)} \\
  $g_{\parallel}$ & $4.412 \pm 0.008$ \\
  $g_{\perp}$ & $1.301 \pm 0.002$ \\
  \colrule
  \multicolumn{2}{l}{Direction of $\mathbf{B}$, from [110]} \\
  rotation towards [100], in (001) & $10^{\circ}$ (best fit $9.7^{\circ} \pm 0.4^{\circ}$) \\
  tilt towards [001] & $5.0^{\circ} \pm 0.1^{\circ}$ \\
  \end{tabular}
  \end{ruledtabular}
\end{table}

With the literature parameters of Table~\ref{tab:eu2pl_params} and no adjustment, the cubic Eu$^{2+}$ spectrum accounts for the strong band (orange diamonds in Fig.~\ref{fig:EuCaF2_Eu2pl_sim}) and for 19 weaker interactions below 0.32~T. Only the direction of $\mathbf{B}$ in the crystal frame is fitted. The six band centers fix its rotation from [110] towards [100] at $9.7^{\circ} \pm 0.4^{\circ}$, but hardly constrain its tilt out of the (001) plane, which is instead set by L1 below.

L1 comprises eight interactions on a line through the origin with $g = 1.3514 \pm 0.0008$, the signature of an effective spin-1/2 Kramers doublet. Of the paramagnetic ions detected by ICPMS in Eu:CaF$_{2}$ (Table~\ref{tab:icpms}), only Nd$^{3+}$ in its tetragonal site, whose $g$ ranges from $g_{\perp} = 1.301$ to $g_{\parallel} = 4.412$~\cite{bleaney1956uranium}, can produce this value. It requires a site whose axis lies $85.0^{\circ}$ from $\mathbf{B}$: the [001] site, with $\mathbf{B}$ tilted $5.0^{\circ} \pm 0.1^{\circ}$ towards [001]. The [100] and [010] sites are then predicted at $g = 3.68$ and $2.74$, crossing the modes between 0.17 and 0.47~T among the dense Eu$^{2+}$ transitions, where no matching interactions have been identified.

The remaining 49 weak interactions match neither ion. Above 0.6~T, 23 of them lie on four straight-line families with similar slopes, A--D (Table~\ref{tab:EuCaF2_g}), each passing through interactions on five to seven modes. We found these by testing every straight line through pairs of these interactions (slopes 13--60~GHz/T) for interactions within 5~mT on at least five modes. With the interactions of each mode shifted at random, the same search finds on average 0.14 such families, and never four in 500 trials. The families have $g_{\mathrm{eff}} = 1.76$--$2.02$ and extrapolate to zero frequency at 0.36--0.63~T, i.e.\ they lie 10--18~GHz above the $g \approx 2$ resonance. This is the behavior of fine-structure transitions of an $S$-state ion ($S \geq 1$) in a non-cubic site with a zero-field splitting of order 10~GHz, such as Eu$^{2+}$ or Gd$^{3+}$ near a charge compensator or Fe$^{3+}$. At 20~mK only the lowest level is populated, so each magnetically inequivalent orientation of such a center contributes essentially one line, and the four families may correspond to different orientations of one or two centers. Literature parameters for the tetragonal and trigonal Gd$^{3+}$ sites~\cite{edgar1975local} and for the sites of Sm$^{3+}$, Dy$^{3+}$ and Er$^{3+}$~\cite{weber1964paramagnetic} do not reproduce them at this orientation. The other 26 interactions, 13 of them below 0.6~T, remain unassigned.

\begin{table}[!tb]
  \caption{\label{tab:EuCaF2_g} Lines through the weak interactions of Fig.~\ref{fig:EuCaF2_Eu2pl_sim} that do not lie on a Eu$^{2+}$ transition: $g$ (or the effective $g_{\mathrm{eff}}$) and zero-field intercept $f_{0}$ from a straight-line fit $f = g \mu_{B} B / h + f_{0}$, and the number $N$ of interactions on each line. L1 is fitted through the origin. Families A--D are straight lines among the interactions above 0.6~T that each pass through interactions on at least five modes.}
  \begin{ruledtabular}
  \begin{tabular}{llccc}
  Line & Assignment & $g$ & $f_{0}$ (GHz) & $N$ \\
  \colrule
  L1 & Nd$^{3+}$, [001] site & $1.3514 \pm 0.0008$ & 0 & 8 \\
  A & unassigned & $2.022 \pm 0.005$ & $-17.81 \pm 0.07$ & 6 \\
  B & unassigned & $1.853 \pm 0.020$ & $-10.96 \pm 0.27$ & 7 \\
  C & unassigned & $1.922 \pm 0.012$ & $-9.58 \pm 0.16$ & 5 \\
  D & unassigned & $1.763 \pm 0.029$ & $-11.33 \pm 0.45$ & 5 \\
  \end{tabular}
  \end{ruledtabular}
\end{table}

\subsubsection{Yb:CaF$_{2}$}

\begin{figure}[!htb]
    \centering
    \includegraphics[width=0.9\linewidth]{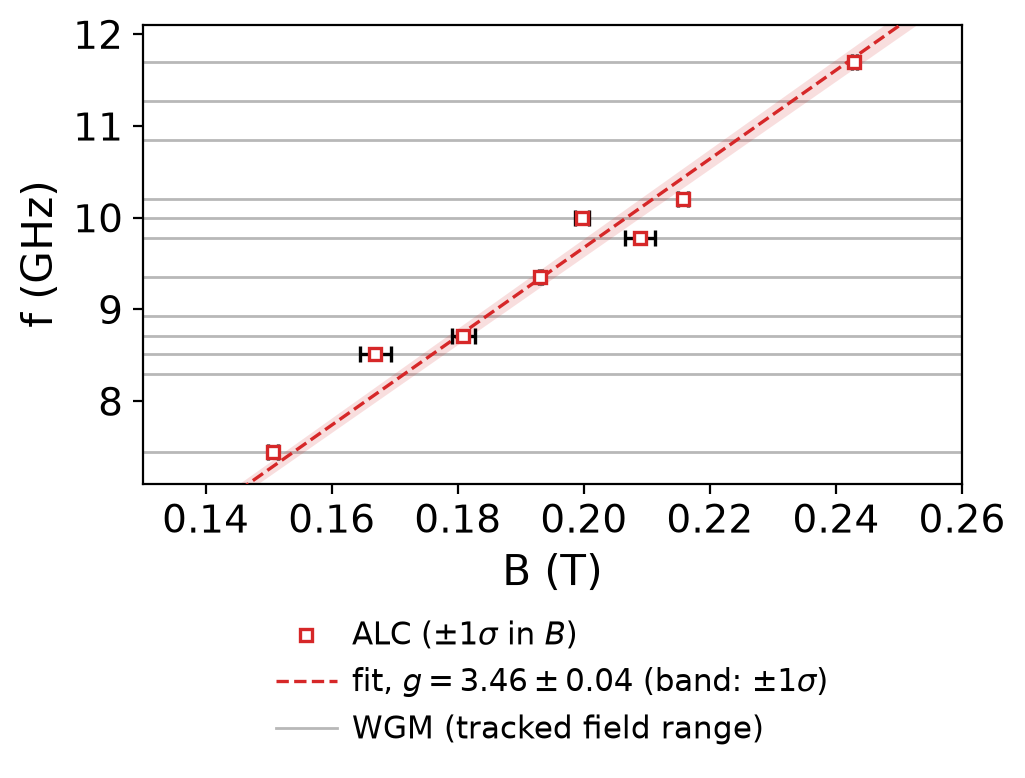}
    \caption{Interaction fields of Yb:CaF$_{2}$ at 20~mK versus WGM frequency. Error bars are $\pm1\sigma$ fit uncertainties in $B$ (smaller than the symbols where not visible). The dashed line is a fit through the origin, $f = g \mu_{B} B / h$, giving $g = 3.46 \pm 0.04$, where the uncertainty is the standard error of the individual interaction $g$ values; the shaded band shows $\pm1\sigma$. Grey lines mark the WGM frequencies.}
    \label{fig:ybcaf2_alc_g-factor}
\end{figure}

The strongly coupled Yb$^{3+}$ ensemble broadens the WGMs of this crystal to half-widths $\kappa$ of 3--11~MHz and makes their frequencies drift with field, so isolated kinks in $f_{0}(B)$ do not reliably mark interactions. Instead, the interaction fit described in Sec.~\ref{sec:experiment} is repeated at 3~mT intervals along each mode, and the local maxima of its significance are kept. A straight line through the origin, $f = g \mu_{B} B / h$, passes through one of these interactions on 8 of the 17 modes, between 7.4 and 11.7~GHz (Fig.~\ref{fig:ybcaf2_alc_g-factor}), and gives $g = 3.46 \pm 0.04$. The quoted uncertainty is the standard error of the $g$ values of the individual interactions; it exceeds the uncertainty propagated from the fitted crossing fields because the interactions scatter about the line by up to 9~mT. The Yb$^{3+}$ feature appears at the fields predicted by this $g$ on all 17 modes, up to 18.6~GHz, but on the remaining modes it shifts $f_{0}$ by much less than $\kappa$ and does not give a resolvable interaction. Weaker features that show no consistent field--frequency dependence across modes are not assigned. Based on this analysis, the ground-state multiplet ($^{2}F_{7/2}$) of Yb$^{3+}$ in a cubic crystal-field environment (T$_c$ symmetry) is identified as the most likely origin of the interactions displayed in Fig.~\ref{fig:ybcaf2_alc_g-factor}. This assignment is supported by the close agreement between the extracted $g$-factor and values reported previously in the literature, as summarized in Table~\ref{tab:YbCaF2_g}.

\begin{table}[!htb]
  \caption{\label{tab:YbCaF2_g} Comparison of measured and published $g$ factors for Yb$^{3+}$ in CaF$_{2}$.}
  \begin{ruledtabular}
  \begin{tabular}{cccc}
  Parameter & Present Work & \cite{10.1007/978-1-4615-7373-9_13, PhysRev.155.279, kirton1969zeeman} & \cite{Falin2004YbCaF2} \\
  \colrule
  $g$ & $3.46 \pm 0.04$ & $3.443$ & 3.476 \\
  \end{tabular}
  \end{ruledtabular}
\end{table}

\FloatBarrier
\section{Conclusion}
\label{sec:conclusion}

We have used whispering-gallery-mode spectroscopy at 20~mK, together with ICPMS, to identify paramagnetic impurities in undoped CaF$_{2}$ and in Th-, Eu- and Yb-doped crystals. Locating each interaction by fitting the field dependence of the mode frequency and linewidth determines its crossing field with a typical uncertainty of 0.1~mT and gives its significance, so that closely spaced and weak features can be separated and compared quantitatively with spin-Hamiltonian predictions.

In Th:CaF$_{2}$ the dominant paramagnetic impurity is Mn$^{2+}$ in cubic sites, identified by its resolved $^{55}$Mn hyperfine lines on seven modes. Its spin-Hamiltonian parameters, $g = 2.0023 \pm 0.0016$ and $A = (-95.2 \pm 0.6) \times 10^{-4}$~cm$^{-1}$, agree with Ref.~\cite{VANGORKOM1970905}, and the near-complete polarization of the electron spins at 20~mK fixes the sign of $A$. Mn was not detected by ICPMS, which places its density below $3.7 \times 10^{15}$~cm$^{-3}$. No paramagnetic center associated with thorium was identified. Thirteen further interactions, all but one below 0.25~T, and the interactions observed at 4~K above 2.4~T remain unassigned; one of the low-field candidate lines has the character of an $S \geq 1$ center with a zero-field splitting of about 11~GHz. In Eu:CaF$_{2}$, cubic Eu$^{2+}$ with literature parameters accounts for the strong band and for 19 weaker interactions below 0.32~T, Nd$^{3+}$ in its tetragonal site for one further line, and four families of unassigned lines indicate $S$-state centers with zero-field splittings of order 10~GHz, such as Eu$^{2+}$ or Gd$^{3+}$ near a charge compensator, or Fe$^{3+}$. In Yb:CaF$_{2}$ the interactions are those of Yb$^{3+}$ in cubic sites, and the same ions dominate the microwave loss, lowering the quality factors at 20~mK, while the $Q$ factors of Th:CaF$_{2}$ improve on cooling.

These results show that microwave cavity spectroscopy at millikelvin temperatures detects paramagnetic impurities below the detection limits of ICPMS, and that mass spectrometry alone does not establish the paramagnetic content of a crystal. For the $^{229}$Th nuclear clock, where defects and impurities of the host contribute to the nuclear linewidth~\cite{li2026origin}, the method identifies the paramagnetic species present and bounds their concentrations in the crystals used. Measurements at a known crystal orientation, or at several orientations, would allow the remaining unassigned lines to be identified through their angular dependence.

\begin{acknowledgments}
This work was supported by the Australian Research Council Centre of Excellence for Dark Matter Particle Physics (CE200100008) and by the Austrian Science Fund (FWF) [Grant DOI: 10.55776/J4834]. M.G.\ is supported by the Australian Research Council Future Fellowship FT250100168.
We acknowledge the facilities of Core Research Infrastructure at The University of Western Australia, a facility funded by the University, State, and Commonwealth Governments of Australia.
\end{acknowledgments}

\section*{Data availability}

The data that support the findings of this article are not publicly available because the raw microwave transmission data sets are large. The data are available from the authors upon reasonable request.

\section*{Conflicts of Interest}

The authors declare no conflict of interest.

\section*{AI Use Disclosure}

Claude Code (Anthropic), an AI coding assistant, was used to write the data-analysis software. S.A.S.\ specified the analysis: what was measured and fitted, the models, the acceptance criteria, and the structure of the code; the AI tool generated the code that implements it, including the mode tracking, the interaction fits, the line searches, the spin-Hamiltonian fits, and the figures produced from them. The AI tool was also used to draft and edit parts of the text. All code was reviewed by S.A.S., and its output was checked against diagnostic plots of every analyzed mode, against literature parameters (Mn$^{2+}$ and Eu$^{2+}$ in CaF$_{2}$, Yb$^{3+}$ $g$ factors), and against randomized controls of the line searches. All scientific decisions, assignments, and conclusions are those of the authors, who take full responsibility for the content.

\appendix

\section{ICPMS results}
\label{app:icpms}

Table~\ref{tab:icpms_raw} lists the concentrations of all 55 measured isotopes in the analyzed solutions, as reported, for each crystal and for the procedural blank, together with the detection limits. Table~\ref{tab:icpms_net} gives the same data after subtraction of the procedural blank, converted to atomic ppm relative to Ca as in Table~\ref{tab:icpms}: a value is given where the blank-subtracted signal exceeds the detection limit, and an upper limit otherwise.

\squeezetable
\begin{table*}[p]
  \caption{\label{tab:icpms_raw} ICPMS concentrations of all measured isotopes in the analyzed solution, in pg/g, as reported for each crystal and for the procedural blank (uncertainties are $1\sigma$ in the last digits). Each crystal was diluted by a factor of about $1.1\times10^{5}$ before analysis. Res.: low (LR) or medium (MR) mass resolution. LOD: detection limit, three times the standard deviation of five instrument blanks.}
  \begin{ruledtabular}
  \begin{tabular}{llccccccc}
  Isotope & Res. & CaF$_2$ & Th:CaF$_2$ & Eu:CaF$_2$ & Yb:CaF$_2$ & Tm:CaF$_2$ & Blank & LOD \\
  \colrule
  $^{23}$Na & MR & $988.27(69)$ & $1031.4(33)$ & $958.53(19)$ & $894.6(17)$ & $818(14)$ & $591.0(66)$ & $718$ \\
  $^{25}$Mg & MR & $317.7(21)$ & $455.5(43)$ & $246.8(53)$ & $482.6(40)$ & $372.6(14)$ & $240.9(24)$ & $194$ \\
  $^{27}$Al & MR & $251.7(25)$ & $159.0(14)$ & $124.2(23)$ & $409.7(31)$ & $157.7(11)$ & $137.96(92)$ & $11$ \\
  $^{31}$P & MR & $2687(30)$ & $277.4(36)$ & $515.2(41)$ & $1062.2(35)$ & $387.9(32)$ & $2421(18)$ & $79$ \\
  $^{32}$S & MR & $4003.0(30)$ & $4057.0(40)$ & $3951(14)$ & $4079(14)$ & $4095(22)$ & $4004(55)$ & $1163$ \\
  $^{35}$Cl & MR & $3.339(28)\times10^{7}$ & $3.729(40)\times10^{7}$ & $3.390(53)\times10^{7}$ & $3.623(87)\times10^{7}$ & $3.796(67)\times10^{7}$ & $3.686(70)\times10^{7}$ & $2.3\times10^{5}$ \\
  $^{39}$K & MR & $463(21)$ & $204(14)$ & $639.3(64)$ & $210.4(87)$ & $297(17)$ & $257(23)$ & $295$ \\
  $^{43}$Ca & MR & $4.561(76)\times10^{6}$ & $4.259(14)\times10^{6}$ & $4.469(23)\times10^{6}$ & $3.2705(72)\times10^{6}$ & $5.57(16)\times10^{6}$ & $2371(33)$ & $1072$ \\
  $^{47}$Ti & MR & $2.94(61)$ & $4.017(67)$ & $2.49(13)$ & $2.18(26)$ & $2.90(28)$ & $1.09(26)$ & $0.59$ \\
  $^{51}$V & MR & $0.305(13)$ & $0.3147(46)$ & $0.261(11)$ & $0.303(37)$ & $0.244(86)$ & $0.0813(65)$ & $0.29$ \\
  $^{52}$Cr & MR & $4.587(71)$ & $3.52(21)$ & $3.323(13)$ & $5.71(24)$ & $3.806(70)$ & $2.633(29)$ & $0.51$ \\
  $^{55}$Mn & MR & $1.845(53)$ & $1.308(46)$ & $1.224(96)$ & $1.81(10)$ & $2.615(77)$ & $0.940(74)$ & $0.87$ \\
  $^{56}$Fe & MR & $17.47(42)$ & $19.04(17)$ & $22.1(39)$ & $21.471(49)$ & $18.826(28)$ & $10.739(31)$ & $3.1$ \\
  $^{59}$Co & MR & $0.2915(66)$ & $0.281(16)$ & $0.329(20)$ & $0.2649(17)$ & $0.251(12)$ & $0.122(13)$ & $0.24$ \\
  $^{60}$Ni & MR & $5.08(16)$ & $4.69(11)$ & $4.60(14)$ & $4.86(26)$ & $4.813(36)$ & $3.95(24)$ & $1.5$ \\
  $^{63}$Cu & MR & $3.02(19)$ & $4.194(83)$ & $1.714(34)$ & $2.0280(80)$ & $2.447(24)$ & $0.686(87)$ & $1.2$ \\
  $^{66}$Zn & MR & $5.73(11)$ & $16.42(57)$ & $6.32(78)$ & $5.58(53)$ & $14.99(23)$ & $3.63(67)$ & $0.59$ \\
  $^{69}$Ga & MR & $7.47(14)$ & $7.92(21)$ & $8.21(41)$ & $7.58(29)$ & $6.17(64)$ & $1.62(15)$ & $9.8$ \\
  $^{77}$Se & LR & $3493.0(30)$ & $3580(120)$ & $3750(48)$ & $3635(93)$ & $3582(85)$ & $3500(160)$ & $521$ \\
  $^{85}$Rb & LR & $0.5697(51)$ & $0.564(16)$ & $0.628(46)$ & $56.76(31)$ & $0.601(14)$ & $0.400(46)$ & $0.29$ \\
  $^{88}$Sr & LR & $12.89(13)$ & $24.14(22)$ & $1432.7(36)$ & $284(10)$ & $108.42(62)$ & $1.816(62)$ & $6.5$ \\
  $^{89}$Y & LR & $1.01(12)$ & $1.104(18)$ & $50.18(41)$ & $3.806(37)$ & $14.40(41)$ & $0.277(30)$ & $1.6$ \\
  $^{95}$Mo & LR & $0.6644(80)$ & $0.714(46)$ & $0.667(97)$ & $0.711(51)$ & $0.57(10)$ & $0.396(51)$ & $0.30$ \\
  $^{103}$Rh & LR & $0.199(14)$ & $0.196(24)$ & $0.260(32)$ & $0.205(23)$ & $0.183(10)$ & $0.09196(28)$ & $0.21$ \\
  $^{105}$Pd & LR & $4.341(80)$ & $3.68(42)$ & $6.46(30)$ & $3.87(13)$ & $4.06(14)$ & $2.8970(80)$ & $2.7$ \\
  $^{111}$Cd & LR & $0.617(34)$ & $0.615(50)$ & $0.671(87)$ & $0.744(87)$ & $0.72(11)$ & $0.544(41)$ & $0.25$ \\
  $^{118}$Sn & LR & $15.1(11)$ & $11.0(10)$ & $7.24(13)$ & $8.08(16)$ & $6.303(69)$ & $5.50(11)$ & $0.73$ \\
  $^{121}$Sb & LR & $0.454(87)$ & $0.481(53)$ & $0.486(74)$ & $0.45(12)$ & $0.549(22)$ & $0.2463(42)$ & $0.40$ \\
  $^{125}$Te & LR & $1.16(11)$ & $0.94(20)$ & $1.16(36)$ & $1.07(28)$ & $0.78(28)$ & $0.671(29)$ & $1.4$ \\
  $^{133}$Cs & LR & $0.3435(36)$ & $0.2910(84)$ & $0.3371(28)$ & $0.3481(96)$ & $0.2942(32)$ & $0.1755(65)$ & $0.22$ \\
  $^{138}$Ba & MR & $1.360(26)$ & $0.993(17)$ & $1490(48)$ & $9.282(79)$ & $17.51(30)$ & $0.259(42)$ & $0.36$ \\
  $^{139}$La & LR & $1.943(92)$ & $2.241(37)$ & $3.636(32)$ & $11.72(13)$ & $16.19(37)$ & $0.849(59)$ & $3.0$ \\
  $^{140}$Ce & LR & $2.15(18)$ & $2.373(84)$ & $4.85(11)$ & $2.72(10)$ & $7.324(29)$ & $0.472(76)$ & $3.9$ \\
  $^{141}$Pr & LR & $0.587(12)$ & $0.6110(32)$ & $1.585(27)$ & $0.752(26)$ & $0.6020(45)$ & $0.140(21)$ & $0.95$ \\
  $^{146}$Nd & LR & $1.326(10)$ & $1.442(88)$ & $8.583(34)$ & $1.630(63)$ & $3.3370(30)$ & $0.287(35)$ & $2.4$ \\
  $^{147}$Sm & LR & $0.37679(11)$ & $0.444(13)$ & $4.198(44)$ & $0.6486(48)$ & $1.925(61)$ & $0.1170(65)$ & $0.86$ \\
  $^{153}$Eu & LR & $0.198(17)$ & $0.209(10)$ & $12200(59)$ & $1.402(84)$ & $1.001(94)$ & $0.157(12)$ & $0.26$ \\
  $^{157}$Gd & LR & $0.383(55)$ & $0.4685(87)$ & $4.880(38)$ & $0.697(32)$ & $0.487(19)$ & $0.1137(42)$ & $0.72$ \\
  $^{159}$Tb & LR & $0.194(19)$ & $0.1913(73)$ & $0.997(41)$ & $0.2352(75)$ & $0.7294(49)$ & $0.0403(31)$ & $0.35$ \\
  $^{163}$Dy & LR & $0.330(36)$ & $0.367(32)$ & $4.307(59)$ & $0.484(18)$ & $0.408(13)$ & $0.069(17)$ & $0.55$ \\
  $^{165}$Ho & LR & $0.1829(46)$ & $0.184(13)$ & $0.907(12)$ & $0.4240(46)$ & $0.5806(42)$ & $0.0341(10)$ & $0.31$ \\
  $^{166}$Er & LR & $0.209(13)$ & $0.256(26)$ & $1.89(10)$ & $0.920(17)$ & $3.55(16)$ & $0.042(19)$ & $0.40$ \\
  $^{169}$Tm & LR & $0.768(34)$ & $0.827(48)$ & $1.290(45)$ & $1.137(15)$ & $1.5191(67)\times10^{5}$ & $3.8(11)$ & $2.2$ \\
  $^{172}$Yb & LR & $0.584(49)$ & $0.664(66)$ & $1.064(17)$ & $1.275(15)\times10^{5}$ & $36.56(99)$ & $1.44(17)$ & $1.4$ \\
  $^{175}$Lu & LR & $0.16437(13)$ & $0.17719(28)$ & $0.59196(95)$ & $10.309(16)$ & $18.14(40)$ & $0.0384(49)$ & $0.28$ \\
  $^{178}$Hf & LR & $1.75(28)$ & $0.601(46)$ & $0.554(53)$ & $0.881(17)$ & $0.4727(84)$ & $0.0392(48)$ & $0.47$ \\
  $^{182}$W & LR & $1.50(13)$ & $14.84(32)$ & $3.193(74)$ & $2.728(76)$ & $3.584(91)$ & $3.0(12)$ & $0.60$ \\
  $^{185}$Re & LR & $0.1509(69)$ & $0.177(10)$ & $0.206(20)$ & $0.1891(53)$ & $2238(28)$ & $0.111(13)$ & $0.30$ \\
  $^{195}$Pt & LR & $52.328(63)$ & $56.4(13)$ & $62.01(20)$ & $64.7(12)$ & $69.0(27)$ & $73.0(14)$ & $27$ \\
  $^{197}$Au & LR & $48.2(22)$ & $37.5(10)$ & $36.9(20)$ & $83.1(19)$ & $57.5(53)$ & $39.4(17)$ & $11$ \\
  $^{202}$Hg & LR & $7.16(94)\times10^{6}$ & $4.12(19)\times10^{6}$ & $3.40(11)\times10^{6}$ & $3.08(12)\times10^{6}$ & $3.12(13)\times10^{6}$ & $3.105(43)\times10^{6}$ & $1.8\times10^{5}$ \\
  $^{205}$Tl & LR & $1.476(36)$ & $1.268(36)$ & $1.166(19)$ & $1.033(61)$ & $1.032(23)$ & $0.824(23)$ & $0.25$ \\
  $^{208}$Pb & LR & $10.3(14)$ & $7.59(48)$ & $6.70(39)$ & $5.66(23)$ & $5.84(11)$ & $4.77(14)$ & $0.46$ \\
  $^{232}$Th & LR & $0.5178(37)$ & $145.7(26)$ & $0.671(17)$ & $2.122(28)$ & $0.4769(87)$ & $0.104(11)$ & $0.92$ \\
  $^{238}$U & LR & $0.1305(69)$ & $0.1408(64)$ & $0.180(16)$ & $0.5580(21)$ & $0.138(10)$ & $0.0306(12)$ & $0.23$ \\
  \end{tabular}
  \end{ruledtabular}
\end{table*}

\squeezetable
\begin{table*}[p]
  \caption{\label{tab:icpms_net} ICPMS results after subtraction of the procedural blank, in atomic ppm relative to Ca (1~ppm $= 2.45\times10^{16}$~cm$^{-3}$; uncertainties are $1\sigma$ in the last digits). Upper limits are the detection limits from Table~\ref{tab:icpms_raw} converted to atomic ppm. Table~\ref{tab:icpms} is an excerpt. Not quantified: $^{a}$acid matrix or instrument memory (Cl, S, Se from the HCl and acids; P from the reagents; Hg, as advised by the laboratory); $^{b}$procedural blank raised by memory of the preceding Tm-doped sample; $^{c}$carry-over from an earlier, much stronger sample; $^{d}$isobaric interference from the dopant (TmO$^{+}$ on $^{185}$Re, Yb$^{2+}$ on $^{85}$Rb and $^{88}$Sr, YbH$^{+}$ on $^{175}$Lu, YO$^{+}$ on $^{105}$Pd); $^{e}$signal falling steadily through the run, including the blank. Ca (the reference) is omitted.}
  \begin{ruledtabular}
  \begin{tabular}{lccccc}
  Isotope & CaF$_2$ & Th:CaF$_2$ & Eu:CaF$_2$ & Yb:CaF$_2$ & Tm:CaF$_2$ \\
  \colrule
  $^{23}$Na & $<275$ & $<294$ & $<280$ & $<383$ & $<225$ \\
  $^{25}$Mg & $<70$ & $83.1(19)$ & $<72$ & $122.0(24)$ & $<57$ \\
  $^{27}$Al & $37.1(11)$ & $7.35(58)$ & $<3.5$ & $123.5(15)$ & $5.28(42)$ \\
  $^{31}$P & --$^{a}$ & --$^{a}$ & --$^{a}$ & --$^{a}$ & --$^{a}$ \\
  $^{32}$S & --$^{a}$ & --$^{a}$ & --$^{a}$ & --$^{a}$ & --$^{a}$ \\
  $^{35}$Cl & --$^{a}$ & --$^{a}$ & --$^{a}$ & --$^{a}$ & --$^{a}$ \\
  $^{39}$K & $<66$ & $<71$ & $87.8(56)$ & $<93$ & $<54$ \\
  $^{47}$Ti & $0.34(12)$ & $0.575(53)$ & $0.261(54)$ & $0.278(94)$ & $0.271(58)$ \\
  $^{51}$V & $<0.050$ & $<0.054$ & $<0.051$ & $<0.070$ & $<0.041$ \\
  $^{52}$Cr & $0.330(14)$ & $0.161(38)$ & $0.1191(55)$ & $0.725(57)$ & $0.162(12)$ \\
  $^{55}$Mn & $0.145(15)$ & $<0.15$ & $<0.14$ & $<0.20$ & $0.220(15)$ \\
  $^{56}$Fe & $1.060(69)$ & $1.399(29)$ & $1.82(63)$ & $2.357(14)$ & $1.043(31)$ \\
  $^{59}$Co & $<0.036$ & $<0.039$ & $<0.037$ & $<0.050$ & $<0.030$ \\
  $^{60}$Ni & $<0.22$ & $<0.24$ & $<0.23$ & $<0.31$ & $<0.18$ \\
  $^{63}$Cu & $0.323(29)$ & $0.520(18)$ & $<0.16$ & $0.259(17)$ & $0.200(12)$ \\
  $^{66}$Zn & $0.282(91)$ & $1.84(13)$ & $0.37(14)$ & $0.36(16)$ & $1.251(86)$ \\
  $^{69}$Ga & $<1.2$ & $<1.3$ & $<1.3$ & $<1.7$ & $<1.0$ \\
  $^{77}$Se & --$^{a}$ & --$^{a}$ & --$^{a}$ & --$^{a}$ & --$^{a}$ \\
  $^{85}$Rb & $<0.030$ & $<0.032$ & $<0.030$ & --$^{d}$ & $<0.024$ \\
  $^{88}$Sr & $1.111(24)$ & $2.399(26)$ & $146.53(85)$ & --$^{d}$ & $8.76(26)$ \\
  $^{89}$Y & $<0.16$ & $<0.17$ & $5.037(49)$ & $0.4868(67)$ & $1.144(47)$ \\
  $^{95}$Mo & $<0.027$ & $0.0312(67)$ & $<0.028$ & $0.0403(92)$ & $<0.022$ \\
  $^{103}$Rh & $<0.018$ & $<0.019$ & $<0.018$ & $<0.025$ & $<0.015$ \\
  $^{105}$Pd & $<0.22$ & $<0.24$ & --$^{d}$ & $<0.31$ & $<0.18$ \\
  $^{111}$Cd & $<0.019$ & $<0.021$ & $<0.020$ & $<0.027$ & $<0.016$ \\
  $^{118}$Sn & --$^{e}$ & --$^{e}$ & --$^{e}$ & --$^{e}$ & --$^{e}$ \\
  $^{121}$Sb & $<0.029$ & $<0.031$ & $<0.029$ & $<0.040$ & $<0.023$ \\
  $^{125}$Te & $<0.095$ & $<0.10$ & $<0.097$ & $<0.13$ & $<0.078$ \\
  $^{133}$Cs & $<0.015$ & $<0.016$ & $<0.015$ & $<0.020$ & $<0.012$ \\
  $^{138}$Ba & $0.0705(34)$ & $0.0503(31)$ & $97.3(31)$ & $0.8058(82)$ & $0.905(31)$ \\
  $^{139}$La & $<0.19$ & $<0.20$ & $<0.19$ & $0.960(13)$ & $0.795(30)$ \\
  $^{140}$Ce & $<0.24$ & $<0.26$ & $0.2805(84)$ & $<0.34$ & $0.352(11)$ \\
  $^{141}$Pr & $<0.060$ & $<0.064$ & $0.0920(22)$ & $<0.083$ & $<0.049$ \\
  $^{146}$Nd & $<0.14$ & $<0.15$ & $0.5161(41)$ & $<0.20$ & $0.1522(48)$ \\
  $^{147}$Sm & $<0.050$ & $<0.054$ & $0.2435(29)$ & $<0.070$ & $0.0866(39)$ \\
  $^{153}$Eu & $<0.015$ & $<0.016$ & $720.4(51)$ & --$^{c}$ & --$^{c}$ \\
  $^{157}$Gd & $<0.040$ & $<0.043$ & $0.2720(26)$ & $<0.056$ & $<0.033$ \\
  $^{159}$Tb & $<0.020$ & $<0.021$ & $0.0540(23)$ & $<0.027$ & $0.03122(95)$ \\
  $^{163}$Dy & $<0.030$ & $<0.032$ & $0.2340(36)$ & $<0.042$ & $<0.024$ \\
  $^{165}$Ho & $<0.017$ & $<0.018$ & $0.04751(68)$ & $0.02899(35)$ & $0.02386(72)$ \\
  $^{166}$Er & $<0.021$ & $<0.022$ & $0.0993(55)$ & $0.0644(19)$ & $0.1510(82)$ \\
  $^{169}$Tm & --$^{b}$ & --$^{b}$ & --$^{b}$ & --$^{b}$ & $6470(190)$ \\
  $^{172}$Yb & $<0.071$ & $<0.076$ & $<0.073$ & $9040(100)$ & --$^{c}$ \\
  $^{175}$Lu & $<0.014$ & $<0.015$ & $0.02839(30)$ & --$^{d}$ & $0.745(27)$ \\
  $^{178}$Hf & $0.084(14)$ & $0.0297(25)$ & $0.0259(27)$ & $0.0579(12)$ & $<0.019$ \\
  $^{182}$W & $<0.029$ & $0.609(63)$ & $<0.029$ & $<0.040$ & $<0.026$ \\
  $^{185}$Re & $<0.014$ & $<0.015$ & $<0.015$ & $<0.020$ & --$^{d}$ \\
  $^{195}$Pt & $<1.2$ & $<1.3$ & $<1.2$ & $<1.7$ & $<0.98$ \\
  $^{197}$Au & $<0.47$ & $<0.51$ & $<0.48$ & $2.72(16)$ & $0.66(21)$ \\
  $^{202}$Hg & --$^{a}$ & --$^{a}$ & --$^{a}$ & --$^{a}$ & --$^{a}$ \\
  $^{205}$Tl & --$^{e}$ & --$^{e}$ & --$^{e}$ & $<0.015$ & $<8.9\times10^{-3}$ \\
  $^{208}$Pb & --$^{e}$ & --$^{e}$ & --$^{e}$ & --$^{e}$ & --$^{e}$ \\
  $^{232}$Th & $<0.035$ & $5.91(11)$ & $<0.036$ & $0.1066(16)$ & $<0.028$ \\
  $^{238}$U & $<8.4\times10^{-3}$ & $<9.0\times10^{-3}$ & $<8.6\times10^{-3}$ & $0.02717(14)$ & $<6.9\times10^{-3}$ \\
  \end{tabular}
  \end{ruledtabular}
\end{table*}

\clearpage

\bibliography{CaF2_impurities_PRB}%

\end{document}